\documentclass[journal]{IEEEtran}

\usepackage{cite}
\usepackage{amsmath}
\usepackage{amssymb,amsthm}
\usepackage[mathscr]{eucal}
\usepackage{accents}

\usepackage{algpseudocode}
\newsavebox{\ieeealgbox}
\newenvironment{boxedalgorithmic}
  {\begin{lrbox}{\ieeealgbox}
   \begin{minipage}{\dimexpr\columnwidth-2\fboxsep-2\fboxrule}
   \begin{algorithmic}[1]}
  {\end{algorithmic}
   \end{minipage}
   \end{lrbox}\noindent\fbox{\usebox{\ieeealgbox}}}
   
\usepackage{cleveref}
\usepackage{color}

\usepackage{graphicx,psfrag,epstopdf}
\ifCLASSOPTIONcompsoc 
\usepackage[caption=false,font=normalsize,labelfont=sf,textfont=sf]{subfig}
\else
\usepackage[caption=false,font=footnotesize]{subfig}
\fi

\newtheorem{theorem}{Theorem}
\newtheorem{corollary}{Corollary}
\newtheorem{lemma}{Lemma}

\def\defeq{{\stackrel{\Delta}{=}}}

\def\eg{{\it e.g., \/}}

\def\ie{{\it i.e.,\ \/}}

\newcommand{\Smsc}{\mathscr{S}}
\def\qr{\mbox{\tiny Q}}
\def\rm{\mbox{\tiny R}}
\def\tra{\intercal}             
               
\DeclareMathOperator{\Tr}{tr}   
\newcommand{\ubar}[1]{\underaccent{\bar}{#1}}

\begin{document}
\title{ONLINE LEARNING AND OPTIMIZATION OF MARKOV JUMP AFFINE MODELS}

\author{Sevi~Baltaoglu,~\IEEEmembership{Member,~IEEE,}
        Lang~Tong,~\IEEEmembership{Fellow,~IEEE,}
        and~Qing~Zhao,~\IEEEmembership{Fellow,~IEEE}% <-this % stops a space
%\thanks{This work was supported by the National Science Foundation under
%Grant CNS-1135844 and CNS-1248079 and by the Army Research Office under Grant W911NF-12-1-0271.
% }% <-this % stops a space
\thanks{Sevi Baltaoglu, Lang Tong, and Qing Zhao are with the School of Electrical and Computer Engineering, Cornell University, Ithaca, NY, 14850 USA e-mail: {\tt \{msb372,lt35,qz16\}@cornell.edu}.}%
\thanks{This work was supported in part by the National Science Foundation under Grants CNS-1135844 and 1549989 and by the Army Research Office under Grant W911NF-12-1-0271.}%
\thanks{Part of the work has been presented at 41st IEEE International Conference on Acoustics, Speech, and Signal Processing, March 2016 and will be presented at 2016 IEEE Workshop on Statistical Signal Processing.}}% <-this % stops a space

% The paper headers (The only time the second header will appear is for the odd numbered pages after the title page when using the twoside option.)



\maketitle

\begin{abstract}
The problem of online learning and optimization of unknown Markov jump affine models is considered. An online learning policy, referred to as Markovian simultaneous perturbations stochastic approximation (MSPSA), is proposed for two different optimization objectives: (i) the quadratic cost minimization of the regulation problem and (ii) the revenue (profit) maximization problem. It is shown that the regret of MSPSA grows at the order of the square root of the learning horizon.  Furthermore, by the use of van Trees inequality, it is shown that the regret of any policy grows no slower than that of MSPSA, making MSPSA an order optimal learning policy. In addition, it is also shown that the MSPSA policy converges to the optimal control input almost surely as well as in the mean square sense. Simulation results are presented to illustrate the regret growth rate of MSPSA and to show that MSPSA can offer significant gain over the greedy certainty equivalent approach.
\end{abstract}

\begin{IEEEkeywords}
Online learning, stochastic approximation, stochastic Cramer-Rao bounds, continuum-armed bandit, sequential decision making.
\end{IEEEkeywords}
% \fi
\IEEEpeerreviewmaketitle

\section{Introduction}
\label{sec:intro}

\IEEEPARstart{W}{e} consider the problem of {\em online learning and optimization} of  affine memoryless models with unknown parameters that follow a Markov jump process.  By online learning and optimization we mean that the control input of the unknown model is chosen sequentially to minimize the expected total cost or to maximize the expected cumulative reward procured over a time horizon $T$. In this context, the online learning problem is one of exploration and exploitation; the need of exploring the space of unknown parameters must be balanced by the need of exploiting the knowledge acquired through learning.

For online learning problems with deterministic unknown parameters, a commonly used performance measure is the so-called {\it regret}\footnote{The notion of regret is made precise in Section 2.}, defined by the difference between the cumulative cost/reward of an online learning policy and that of a decision maker who knows the model completely and sets the input optimally. The regret grows monotonically with the time horizon $T$, and the rate of growth measures the efficiency of online learning policies. 

The online learning problem considered in this paper is particularly relevant in dynamic pricing problems when the consumers' demand is unknown and possibly varying stochastically \cite{Jia&Tong&Zhao:13Allerton, Bertsimas&etal:06, Lobo&Boyd:03Informs, Keskin&Zeevi:14}. The goal of dynamic pricing is to set the price sequentially, using the observations from the previous sales, to match a certain contracted demand. Besides applications in dynamic pricing, results are also relevant to the learning and control problem of Markov jump linear systems with unknown parameters \cite{ Logothetis&Krishnamurthy:TSP99, Doucet&etal:TSP01, Costa&Fragoso&Marques:05Book}. 

In this paper, we study the online learning and optimization problem of Markov jump affine models under two different objectives: (i) target matching with a quadratic cost and (ii) revenue (profit) maximization. Our goal is to establish fundamental limits on the rate of regret growth for Markov jump affine models and develop an online learning policy that achieves the lowest possible regret growth.

\subsection{Related Work}
\label{ssec:relatedwork}

Without Markov jump as part of the model, \ie when there is a single state, the problem considered here is the classical problem of control in experiment design studied by Anderson and Taylor \cite{Anderson&Taylor:76}. Anderson and Taylor proposed a certainty equivalence rule where the input is determined by using the maximum likelihood estimates of system parameters as if they were the true parameters. Despite its intuitive appeal, the Anderson-Taylor rule was  shown to be suboptimal for the quadratic regulation problem by Lai and Robbins in \cite{Lai&Robbins:AAM82} and also for the revenue maximization problem by den Boer and Zwart in \cite{Boer&Zwart:14}. In fact, there is a non-zero probability that the Anderson-Taylor rule produces an input 	which converges to a suboptimal value for both cases; therefore, this rule results in {\it incomplete learning} and a linear growth of regret.

For the scalar model in which the quadratic cost of the regulation problem is to be minimized, Lai and Robbins \cite{Lai&Robbins:79AS} showed that a Robbins-Monro stochastic approximation approach achieves the optimal regret order of $\Theta(\log T)$. Later, Lai and Wei \cite{Lai&Wei:87} showed that this regret order is also achievable for a more general linear dynamic system by an adaptive regulator that uses least square estimates of a reparametrized model and ensures convergence via occasional uses of white-noise probing inputs. The result was further generalized by Lai \cite{Lai:86} to multivariate linear dynamic systems with a square invertible system matrix. The result presented in this paper can be viewed as a generalization of this line of work to allow both time-varying linear models and time-invariant models with a non-invertible system matrix.
 
The problem considered in this paper also falls into the category of continuum-armed bandit problem where the control input is chosen from a subset of $\Re^n$ with the goal of minimizing expected cost (or maximizing expected reward) that is an unknown continuous function of the input. This problem was introduced by Agrawal \cite{Agrawal:95SIAM} who studied the scalar problem and proposed a policy that combines certainty equivalence control with Kernel estimator-based learning. Agrawal showed that this policy has a regret growth rate of $O(T^{3/4})$ for a uniformly Lipschitz expected cost function. Later, Kleinberg \cite{Kleinberg:04} proved that the optimal growth rate of regret for this problem cannot be smaller than $\Omega(T^{2/3})$ and proposed a policy that achieves $O(T^{2/3}(\log(T))^{1/3})$. Kleinberg \cite{Kleinberg:04} also considered the multivariate problem , \ie $n>1$, and showed that an adaptation of Zinkevich's greedy projection algorithm achieves the regret growth rate of $O(T^{3/4})$  if the cost function is smooth and convex on a closed bounded convex input set. 

Within the continuum-armed bandit formulation, the work of Cope \cite{Cope:09} is particularly relevant because of its use of stochastic approximation to achieve the order-optimal regret growth of $\Omega(\sqrt{T})$ for a different class of cost functions. Cope's results (both the regret lower bound and the Kiefer-Wolfowitz technique), unfortunately, cannot be applied here because of the time-varying Markov jump affine models treated here. Also relevant is the work of Rusmevichientong and Tsitsiklis \cite{Rusmevichientong&Tsitsiklis:10} on the so-called linearly parameterized bandit problem where the objective is to minimize a linear cost with input selected from the unit sphere.  A learning policy developed in \cite{Rusmevichientong&Tsitsiklis:10} is shown to achieve the lower bound of $\Omega(\sqrt{T})$ using decoupled exploration and exploitation phases. Even though the model considered in this paper is similar to the one in \cite{Rusmevichientong&Tsitsiklis:10} in terms of the observed output being a linear function of the input, in our paper, the unknown model parameters follow a Markov jump process and the specific cost functions studied are quadratic; thus the problem objective is different.

There is a considerable amount of work on dynamic pricing problem  with the objective of revenue maximization under a demand model uncertainty in different areas such as operations research, statistics, mathematics, and computer science. In  \cite{Kleinberg&Leighton:03}, a multi-armed bandit approach with a regret growth rate of $O(\sqrt{T\log{T}})$ was proposed for a nonparametric formulation of the problem. See also \cite{Broder&Rusmevichientong:12} where the same problem under a general parametric demand model is considered and a modified version of myopic maximum likelihood based policy is shown to achieve the regret order of $O(\sqrt{T})$, and \cite{Boer&Zwart:14} where a similar result is obtained for a class of parametric demand models. In both \cite{Kleinberg&Leighton:03} and \cite{Broder&Rusmevichientong:12}, authors proved that the lower bound for regret growth rate is  $\Omega(\sqrt{T})$.

Besides more general classes of demand models, affine model similar to the one in this paper has been also studied extensively; \eg \cite{Bertsimas&etal:06,Lobo&Boyd:03Informs,Keskin&Zeevi:14}. In both \cite{Bertsimas&etal:06} and \cite{Lobo&Boyd:03Informs}, it is shown that approximate dynamic programming solutions may outperform greedy method numerically. A special case of our formulation of revenue maximization problem without any Markov jump characteristics (with time-invariant model parameters) is previously investigated by Keskin and Zeevi \cite{Keskin&Zeevi:14}. Keskin and Zeevi proposed a semi-myopic policy that uses orthogonal pricing idea to explore and learn the system. They showed that the lowest possible regret order is $\Omega(\sqrt{T})$ for any policy, and their semi-myopic policy achieves this bound up to a logarithmic factor; \ie $O(\sqrt{T}\log{T})$.

Even though the system model is assumed to be time-invariant in most of the literature, there is a considerable amount of work especially in dynamic pricing that deals with time-varying demand models due to unpredictable environmental factors affecting demand; \eg see \cite{Aviv&Pazgal:05} for a demand model that evolves according to a discrete state space Markov chain in a revenue management with finite inventory problem, and \cite{Balvers&Casimano:90} for a dynamic programming formulation of a profit maximization problem with an unknown demand parameter following an autoregressive process. See also \cite{Keskin&Zeevi:13} for a revenue maximization problem with an affine demand model where the model parameters are time-varying, yet the cumulative change in the model parameters over the time horizon $T$ is bounded. Since Keskin and Zeevi \cite{Keskin&Zeevi:13} measure the regret of a policy by the difference between the cumulative cost of the policy and that of a clairvoyant who knows all the future temporal changes exactly and chooses the optimal action, their characterization of regret is too pessimistic for the Markov jump model considered here. 

Some other examples of related work on online learning with time-varying models apart from dynamic pricing are \cite{Besbes&etal:13} and \cite{Yin&Ion&Krishnamurthy:09MP}. In \cite{Besbes&etal:13},  Besbes, Gur, and Zeevi studied the online learning problem of more general time-varying cost functions where the cumulative temporal changes is restricted to a budget similar to \cite{Keskin&Zeevi:13}. However, their characterization of regret is also similar to \cite{Keskin&Zeevi:13} and thus, incomparable with the one in this paper. Yin, Ion, and Krishnamurthy \cite{Yin&Ion&Krishnamurthy:09MP} also considered the problem of estimating a randomly evolving optimum of a cost function which follows a Markov jump process. Their analysis deals with the convergence of the estimate obtained via stochastic approximation to the limit (stationary) solution, whereas in this paper, we are concerned about estimating the optimum of the cost function at each time instant given the previous state of the Markov chain. Moreover, different than our paper, their analysis relies on the availability of the noisy observations of the cost function gradient and they do not characterize regret.

\subsection{Summary of Results}
\label{ssec:contributions}

The main contribution of this paper is the generalization of online learning of time-invariant affine models to that of Markov jump affine models. Extending Spall's stochastic approximation method \cite{Spall:92} to the optimization problem of an objective function that evolves according to a Markov jump process, we propose an online learning policy, referred to as Markovian simultaneous perturbations stochastic approximation (MSPSA). We show that MSPSA achieves the optimal regret order of $\Theta(\sqrt{T})$ for two different objective functions studied for the affine model: (i) quadratic regulation and (ii) the revenue maximization. Furthermore, we also show that the control input of MSPSA policy converges to the optimal solution both with probability one and in mean square as $T \rightarrow \infty$. Therefore, the proposed policy eventually learns the optimal solution.

A key implication of our results is that, in comparing with Lai's result on the learning problem of a time-invariant affine model with the quadratic regulation objective \cite{Lai:86}, modulating a linear model by a Markov jump process introduces substantial learning complexity; hence, the regret order increases from $\Theta(\log T)$ to $\Theta(\sqrt{T})$. As a special case, we also show that, even in the absence of Markovian jump, when the system matrix is full column rank but not invertible, the best regret order is also $\Theta(\sqrt{T})$.  It worths noting that adding just one row to a square and invertible matrix can change the worst case regret from $\Theta(\log T)$ to $\Theta(\sqrt{T})$.

In the second part of this paper, we study the profit maximization problem with Markov jump demand. To this end, we generalize the lower bound obtained by Keskin and Zeevi \cite{Keskin&Zeevi:14} for time-invariant demand to Markov jump demand, and we show that this bound is achievable by the MSPSA policy for a more general case with Markov jump characteristics. 

The results presented here are obtained using several techniques developed in different contexts.  The MSPSA policy is a generalization of Spall's stochastic approximation method to the optimization problem of an objective function following a Markov jump process. To show the optimality of MSPSA, we use the van Trees inequality \cite{Gill&etal:95}  to lower bound the estimation error, which is the technique used in \cite{Keskin&Zeevi:14}. Lastly, a result on the convergence of non-negative almost supermartingales \cite{Robbins:1985} is used to obtain the convergence result for MSPSA policy.

\section{Problem Formulation}

The model considered here is an affine model, modulated by an exogenous finite state time-homogeneous Markov chain $(\Smsc,P)$ where $\Smsc=\{1,\cdots, K\}$ is the state space and $P=[p_{i,j}]$ the transition probability matrix. We assume that the state space $\Smsc$ and the transition matrix $P$ are unknown.

Each state $k \in \Smsc$  of the Markov chain is associated with an affine model whose parameters are denoted by $\theta_k=(A_k,b_k)$ where $A_k \in \Re^{m \times n}$ has full column rank and $b_k \in \Re^m$. All system parameters  $\theta=\{\theta_k\}_{k=1}^K$ are assumed deterministic and unknown. At time $t$, the input-output relation of the system is given by
\begin{IEEEeqnarray}{rCl}
\label{eq:linear}
y_t & = & A_{s_t}x_t + b_{s_t} + w_t,
\end{IEEEeqnarray}
where $x_t \in \Re^n$ is the control input, $y_t\in \Re^m$ the observable output, $s_t \in \Smsc$ the state of the system, and $w_t \in \Re^m$ is a random vector that captures the system noise. It is assumed that the random noise $w_t$ is drawn from a possibly state dependent distribution $f_{s_t}(.)$ with zero mean (without loss of generality) and unknown finite variance $\Sigma_w^{(s_t)}$. Furthermore, for any $t \neq t'$, $w_t$ and $w_{t'}$ are conditionally independent given the states $s_t$ and $s_{t'}$.

The objective of {\em online learning and optimization} is to find a control input sequence $\{x_t\}_{t=1}^T$ that minimizes the expected cumulative cost incurred at each stage. Because $x_t$ is determined before $s_t$ is realized, the stage cost $ \mathcal{J}(s_{t-1},x_t)$ at $t$ is a function of $x_t$ and $s_{t-1}$, and the expected cumulative cost is
\begin{displaymath}
 \mathbb{E} \left( \sum_{t=1}^T \mathcal{J}(s_{t-1},x_t) \right),
\end{displaymath}
where $T$ is the learning and optimization horizon. Note that the above quantity is a function of the deterministic parameters $\theta$ and the distributions $P$ and $\{f_i\}_{i=1}^K$.  

For a decision maker who wants to minimize its expected cumulative cost, the difficulty in finding the optimal control input sequence is that the system parameter $\theta$ and the transition matrix $P$ are unknown. If the system parameters $(\theta,P)$ were known, then the decision maker would have used this information along with its observation history to determine the optimal decision rule that minimizes the expected cumulative cost. In that case, the problem could be formulated as a dynamic program and solved via backward induction. We refer to this optimal solution under known model as the optimal input and denote it by $\{x_t^*\}_{t=1}^T$ (which is made precise in the following sections). We assume that a convex compact set $\Pi \subset \Re^n$ containing the optimal input is known to the decision maker. 

Before choosing the control input of period $t$, the only information the decision maker has is a vector $I_{t-1}$ containing its decision and observation history up to time $t-1$, which includes input vector $X^{t-1}=(x_1,\cdots, x_{t-1})$, output vector $Y^{t-1}=(y_1,\cdots, y_{t-1})$, state vector $S^{t-1}=(s_0,\cdots, s_{t-1})$, and the set $\Pi$. Consequently, a policy $\mu$ of a decision maker is defined as a sequence of decision rules, \ie $\mu =(\mu_0, \mu_1, \cdots,\mu_{T-1})$, such that, at time $t-1$, $\mu_{t-1}$ maps the information history vector $I_{t-1}$ to the system input $x_{t}$ at time $t$. We denote the input determined by policy $\mu$ as $x_t^{\mu}$.

To measure the performance of an online learning policy, we use the regret measure as a proxy. In particular, the (cumulative) regret $\mathcal{R}^\mu_T(\theta,P)$ of a learning policy $\mu$ is measured by the difference between the expected cumulative cost of the decision maker, who follows the policy $\mu$, and that of a decision maker who knows the system parameters $(\theta,P)$ and sets the system input optimally, \ie 

\begin{IEEEeqnarray}{rCl}
\label{eq:defregret}
\mathcal{R}_T^{\mu}(\theta,P) & = & \mathbb{E} \left(\sum_{t=1}^T  \mathcal{J}(s_{t-1},x_t^\mu) - \sum_{t=1}^T  \mathcal{J}(s_{t-1},x_{t}^*) \right).
\end{IEEEeqnarray}

Since the regret defined above is a function of system parameters, we characterize the performance of $\mu$ by the worst case regret 
\begin{displaymath}
\bar{\mathcal{R}}_T^\mu \defeq \sup_{\theta,P,\{f_{i}\}_{i=1}^K}  \mathcal{R}^\mu_T(\theta,P).
\end{displaymath}
In terms of the worst-case analysis, it is assumed that the worst-case system parameter $\theta$ is chosen from a compact set $\Theta \subset \Re^{K \times m \times (n+1)}$ for any fixed values of the state space size $K$ and system dimensions $m$ and $n$. Since $\Theta$ is compact, for any $\theta \in \Theta$ and for all $k \in \Smsc$, the largest singular value of $A_k$ is bounded by a positive constant $\bar{\sigma}$, and the parameter $b_{k}$ is bounded by a positive constant $\bar{b}$, \ie $\|b_{k}\|_2 \leq \bar{b}$. It is also assumed that the variance of $w_t$ is bounded, \ie $\mathbb{E}(w_{t,i}^2)\leq \sigma_w^2$ for some positive constant  $\sigma_w$ where $w_{t,i}$ denotes the $i$th entry of $w_t$. The optimal input for the worst-case parameters $(\theta,P)$ certainly has to be contained in the set $\Pi$, and $A_k$ has to be full column rank for every $k \in \Smsc$ for the worst-case parameter $\theta$. Note that  $\bar{\mathcal{R}}_T^\mu $ grows monotonically with $T$.  We are  interested in the learning rule that has the slowest regret growth.

In the following sections, we focus on two different stage costs, hence two different objective functions. The first is a quadratic cost that arises naturally from the regulation problem. In particular, the stage cost at time $t$  is given by
  \begin{IEEEeqnarray}{rCl}
  \label{eq:cost}
   \mathcal{J}^{\qr}(s_{t-1},x_t) & \defeq & \mathbb{E} \left(||y^* - y_t ||_2^2 | s_{t-1}, x_t \right),
  \end{IEEEeqnarray}
where $y^* \in \Re^m$ is a constant target value for output. For the quadratic regulation problem, we assume that the forth order moment of $w_t$ is bounded, \ie $\mathbb{E}(w_{t,i}^4)\leq \sigma_w^4$, in addition to the previous boundedness assumptions.
  
The second stage cost we consider is the minus revenue that arises from revenue maximization problem (or profit maximization problem since profit can be expressed as revenue minus total cost). Specifically, the stage cost of period $t$ is given by
  \begin{IEEEeqnarray}{rCl}
  \label{eq:cost_revenue}
   \mathcal{J}^{\rm}(s_{t-1},x_t) & \defeq & \mathbb{E}  \left( - x_t \cdot y_t | s_{t-1}, x_t \right).
  \end{IEEEeqnarray}
Here, the revenue is calculated as the inner product of the input and the output vector where the entries of the input and the output vector corresponds to the price and the demand of each product, respectively. Therefore, the input and the output dimensions match, $\ie$ $m=n$, for the revenue maximization problem. For this objective, it is assumed that the matrix $A_j$ is negative definite for all $j \in \Smsc$ which is a reasonable assumption in dynamic pricing problems, \eg see \cite{Jia&Tong&Zhao:13Allerton, Keskin&Zeevi:14}. 

\section{Online Learning for Quadratic Regulation}
\label{sec:preliminaries_qr}

In this section, we study the online learning problem with the quadratic cost. We first derive an expression for regret using the optimal solution under known model referred to as the optimal input. Then, we introduce an online learning approach and establish its order optimality via the analysis of its regret growth rate and the analysis of the minimum regret growth rate achievable by any policy. We also show that the input of the online learning policy converges to the optimal input both almost surely and in mean square.

\subsection{Optimal Solution Under Known Model and Regret}  

In order to calculate the regret for any policy, we begin by deriving the optimal solution of a decision maker who knows the system $(\theta,P)$ in addition to $I_{t-1}$ and aims to minimize the expected cumulative cost, \ie
\begin{IEEEeqnarray}{rCl}
\label{eq:cum_risk}
 \min_{\{x_t\}_{t=1}^T} \mathbb{E} \left( \sum_{t=1}^T \mathcal{J}^{\qr}(s_{t-1},x_t) \right).
\end{IEEEeqnarray}
The problem under known model becomes a dynamic program due to the known $(\theta,P)$. Since the Markov process is exogenous, \ie independent of the decision policy, the optimization problem decouples to choosing the system input $x_t$ separately for each decision stage with stage cost given in (\ref{eq:cost}) which is equivalent to
\begin{IEEEeqnarray}{rCl}
\label{eq:risk}
\mathcal{J}^{\qr}(s_{t-1},x_t) & = & \sum_{j} p_{{s_{t-1}},j} \left( \|y^*-A_jx_t-b_j\|_2^2 + \Tr(\Sigma_w^{(j)}) \right) \IEEEnonumber\\
\end{IEEEeqnarray}
by (\ref{eq:linear}). The optimal input $x_t^*$ minimizing the stage cost is then given by 
\begin{IEEEeqnarray}{rCl}
\label{eq:optprice}
x_{s_{t-1}}^* & = & \Bigl(\sum_j p_{{s_{t-1}},j}A_j^{\tra} A_j\Bigr)^{-1}\Bigl(\sum_j p_{{s_{t-1}},j}A_j^{\tra}(y^*-b_j)\Bigr).
\end{IEEEeqnarray}
Thus, the optimal input $x_t^* \in \Pi$ at any time $t$ depends only on the system parameter $\theta$, the transition matrix $P$, and the previous state $s_{t-1}$. In the sequel, we use $x_{s_{t-1}}^*$ to represent $x_t^*$, dropping the explicit parameter dependency on $(\theta,P)$ in the notation.

Hence, the stage regret at $t$, which is the expected difference of the stage cost obtained by policy $\mu$ and the stage cost of the optimal input $x^*_{s_{t-1}}$, can be written as
\begin{IEEEeqnarray}{rCl}
r_t^{\mu}(\theta,P) &=& \mathbb{E} \left(\mathcal{J}^{\qr}(s_{t-1},x_t^\mu) - \mathcal{J}^{\qr}(s_{t-1},x_{s_{t-1}}^*) \right) \IEEEnonumber\\
&=& \mathbb{E} \left(\|A_{s_t}(x_t^{\mu}-x_{s_{t-1}}^*)\|_2^2 \right), \IEEEnonumber
\end{IEEEeqnarray}
which is obtained using the first order optimality condition (FOC) for $x_{s_{t-1}}^*$. The T-period regret given in (\ref{eq:defregret}) can then be expressed as
\begin{IEEEeqnarray}{rCl}
\label{eq:regret}
\mathcal{R}_T^{\mu}(\theta,P) & = & \mathbb{E} \left( \sum_{t=1}^T \|A_{s_t}(x_t^{\mu}-x_{s_{t-1}}^*)\|_2^2 \right).
\end{IEEEeqnarray}

\subsection{MSPSA: An Online Learning Policy}
\label{sec:algorithm}

Here, we present an online learning policy to the quadratic regulation problem that achieves the slowest regret growth rate possible. Referred to as MSPSA, the policy is an extension of the simultaneous perturbation stochastic approximation (SPSA) algorithm proposed by Spall \cite{Spall:92} to Markov jump models considered here. 

Spall's SPSA is a stochastic approximation algorithm that updates the estimate of the optimal input by a stochastic approximation of the objective gradient. The key step is to generate two consecutive observations corresponding to two inputs, that are set to be the current optimal-input estimate perturbed by some random vector in opposite directions, and use them to construct the gradient estimate. In applying this idea to the optimization problem of a Markov jump system, a complication arises due to the uncertainty associated with the system state at the time when the system input is determined; consecutive observations that are used to determine the gradient estimate may correspond to different system states. 

\begin{figure}
  \begin{boxedalgorithmic}
    \For{t = 1 to T}
   	  \If{$s_{t-1}=i$ is observed}
   	  \If{state $i$ is observed for the first time}
   	  \State Let $\hat{x}_{i,1} \in \Pi$ be an arbitrary vector
   	  \State $t_i \gets 0$
   	  \State $e_i \gets 0$
   	  \EndIf
		\If{$e_i = 0$}
		  \State $t_i \gets t_i+1$
		  \State $x_{t} \gets \hat{x}_{i,t_i}+c_{t_i}\Delta_{t_i}$ 
		  
		  where $c_{t_i}=\gamma'_i/(N'_i+t_i)^{0.25}$ with some positive constant $\gamma'_i$ and a non-negative integer $N'_i$, and $\Delta_{t_i}=[\Delta_{t_i,1},...,\Delta_{t_i,n}]^{\tra}$ with $\Delta_{t_i,j}$'s drawn from an independent and identical distribution that is symmetrical around zero, and satisfies 	$|\Delta_{t_i,j}| \leq \xi_1$ and	$\mathbb{E}(1/\Delta_{t_i,j}^2)\leq \xi_2$  for some positive constants $\xi_1$ and $\xi_2$.
		  \State $d_{i,t_i}^{+} \gets \|y_t-y^*\|_2^2$
		  \State $e_i \gets 1$
		\Else
		  \State $x_{t} \gets \hat{x}_{i,t_i}-c_{t_i}\Delta_{t_i}$
		  \State $d_{i,t_i}^{-} \gets \|y_t-y^*\|_2^2$ 
		  \State $e_i \gets 0$
		  \State Update:
			\begin{IEEEeqnarray}{rCl}
			\label{eq:MSPSAupdate_regulation}
				\hat{x}_{i,t_i+1} & \gets & \left(\hat{x}_{i,t_i} - a_{t_i}\left(\frac{d_{i,t_i}^{+}-d_{i,t_i}^{-}}{c_{t_i}}\right)\bar{\Delta}_{t_i}\right)_{\Pi}
			\end{IEEEeqnarray}
			where $(.)_{\Pi}$ denotes the euclidean projection operator onto $\Pi$, $\bar{\Delta}_{t_i} = [1/\Delta_{t_i,1},...,1/\Delta_{t_i,n}]^{\tra}$, and $a_{t_i}=\gamma_i/(N_i+t_i)$ with some positive constant $\gamma_i$ and a non-negative integer $N_i$
		\EndIf
	  \EndIf 
    \EndFor
  \end{boxedalgorithmic}
\caption{MSPSA for Quadratic Regulation} \label{mspsa_for_qr}
\end{figure}

The key idea of MSPSA is to keep track of each state $i \in \Smsc$ and the estimate of the optimal input associated with each state $i \in \Smsc$.  When state $i$ is realized, the estimate of the optimal input associated with state $i$ is perturbed by some random vector and this randomly perturbed estimate is used as input for the next stage. The estimate of the optimal input associated with state $i$ is updated only when we obtain two observations of the system output corresponding to two inputs that are generated by perturbing the current estimate in opposite directions by the same amount right after observing state $i$. 
 
 Details of this implementation is given in \figurename~\ref{mspsa_for_qr}. Whenever a new state $i \in \Smsc$ is observed that has not been observed before, MSPSA policy assigns an arbitrary predetermined vector $\hat{x}_{i,1} \in \Pi$ as the initial estimate of the optimal input $x_i^*$ (line 3-7 of \figurename~\ref{mspsa_for_qr}). At the beginning of each stage $t$, MSPSA checks the previous state $s_{t-1}$ (line 2), and whether any observation is taken using the most recent optimal-input estimate $\hat{x}_{s_{t-1},t_{s_{t-1}}}$ (line 8) where $t_{s_{t-1}}$ is the number of times the optimal-input estimate $\hat{x}_{s_{t-1},t_{s_{t-1}}}$ is updated up to time t (Since two observations, that are taken right after observing state $s_{t-1}$, are used for each update of $\hat{x}_{s_{t-1},t_{s_{t-1}}}$, $t_{s_{t-1}}$ is approximately half of the number of times state $s_{t-1}$ is observed up to $t$.). If an observation has not taken using the most recent estimate yet, the input for that stage is set to be a randomly perturbed $\hat{x}_{s_{t-1},t_{s_{t-1}}}$ (line 10). Otherwise MSPSA sets the input by perturbing the estimate $\hat{x}_{s_{t-1},t_{s_{t-1}}}$ in the opposite direction by the same amount as the previous one  (line 14).  Then, it updates the optimal-input estimate  by a stochastic approximation (line 17) obtained using the stage costs calculated from both observations (line 11 and 15) and projects it onto $\Pi$. The constant $\gamma'_i$ of the perturbation gain sequence $c_{t_i}$ should be chosen larger in the high noise setting for an accurate gradient estimate. The choice of the sequence $a_{t_i}$ used for the update step determines the step size. The non-negative integers $N_i$ of $a_{t_i}$ and $N'_i$ of $c_{t_i}$ can be set to zero as default, but if the update of the optimal-input estimate fluctuates between the borders of $\Pi$ at the beginning of the MSPSA policy, setting $N_i$ greater than zero can prevent this fluctuation. 

\subsection{Regret Analysis for MSPSA}
\label{ssec:upper}

We now analyze the regret performance of MSPSA. Let $\lambda_{min}(.)$ denote the minimum eigenvalue operator, and $e_{i,t_i}=\mathbb{E}\left( \|\hat{x}_{i,t_i}-x_{i}^*\|^2_2|i,t_i\right)$ be the mean squared error (MSE) between the optimal input $x^*_i$ and its estimate $\hat{x}_{i,t_i}$ given state $i$ and  $t_i$, where $t_i$, as defined in previous section, is the number of times the estimate $\hat{x}_{i,t_i}$ has been updated up to time $t$ by MSPSA. The following lemma provides a bound for the decreasing rate of $e_{i,t_i}$, and hence the convergence rate of the estimate to its true value in terms of the number of times the estimate is updated and thus in terms of the number of times the state $i$ has occurred up to $t$ (which is equal either to $2t_i$ or to $2t_i-1$). It shows that the MSE converges to zero with a rate equal or faster than the inverse of the square root of  the number of times state $i$ has occurred. 

\begin{lemma}
 \label{lemma:regret}
For any $i \in \Smsc$, if $\gamma_i \geq 1/(8\lambda_{min}(\sum_j p_{i,j}A_j^{\tra}A_j))$ then there exists a constant $C_i>0$ satisfying $e_{i,t_i} \leq C_i/\sqrt{t_i}$ for any $(\theta,P,\{f_i\}_{i=1}^K)$.
\end{lemma}
\begin{IEEEproof}
See Appendix.
\end{IEEEproof}

To satisfy the condition of Lemma \ref{lemma:regret}, the decision maker, who follows MSPSA, needs to have some information about a lower bound on the minimum eigenvalue of $\sum_j p_{i,j}A_j^{\tra}A_j$, \eg knowing a non-trivial lower bound $\ubar{\sigma}$ for the singular values of the system matrices. This assumption may not be restrictive in practice since the decision maker can set $\gamma_i$ sufficiently large.    

Let the worst-case cumulative input-MSE $\bar{\mathcal{E}}_T^{\mu}$ be the worst-case cumulative MSE between the input $x_t^{\mu}$ of policy $\mu$ and the optimal input $x^*_{s_{t-1}}$, \ie
\begin{displaymath}
\bar{\mathcal{E}}_T^{\mu}  \defeq \sup_{\theta,P,\{f_{i}\}_{i=1}^K}  \mathbb{E} \left( \sum_{t=1}^T \|x_t^{\mu}-x_{s_{t-1}}^*\|_2^2 \right).
\end{displaymath}
Using the result of Lemma $\ref{lemma:regret}$, we provide a bound for the growth rate of the worst-case cumulative input-MSE and the worst-case regret  of MSPSA. Theorem~\ref{thm:upper} shows that the MSPSA policy achieves the regret growth rate of $O(\sqrt{T})$.

\begin{theorem}
 \label{thm:upper}
If $\gamma_i \geq 1/(8\lambda_{min}(\sum_j p_{i,j}A_j^{\tra}A_j))$ for every $i \in \Smsc$, then there exist some positive constants $C$ and $C'$ such that
\begin{displaymath}
\bar{\mathcal{E}}^{\mbox{\tiny MSPSA}}_T \leq C\sqrt{T},
\end{displaymath}
and
\begin{displaymath}
\bar{\mathcal{R}}^{\mbox{\tiny MSPSA}}_T \leq C'\sqrt{T}.
\end{displaymath}
\end{theorem}	
\begin{IEEEproof}
 The input of MSPSA $x_t^{\mbox{\tiny MSPSA}}$ is equal to either $\hat{x}_{i,t_i}+c_{t_i} \Delta_{t_i}$ or $\hat{x}_{i,t_i}-c_{t_i} \Delta_{t_i}$ given $s_{t-1}=i$ and $t_i$. By Lemma~\ref{lemma:regret}, observe that
\begin{IEEEeqnarray}{rCl}
\label{eq:r_i_t_i}
r_{i,t_i} & = & \mathbb{E} \left(\|\hat{x}_{i,t_i} \pm c_{t_i} \Delta_{t_i} -x_i^*\|_2^2|i,t_i\right) \IEEEnonumber \\
& = & e_{i,t_i} + c_{t_i}^2  \mathbb{E} \big(\Delta_{t_i}^{\tra}\Delta_{t_i}\big) \leq  C'_i/\sqrt{t_i}
\end{IEEEeqnarray}
where $C'_i=C_i+(\gamma_i')^2n\xi_1^2$. Let $T_i$ be the number of times the estimate of the optimal input associated with state $i$ has been updated until period $T$. Because MSPSA uses two observations per update, we can express the worst-case cumulative input-MSE for MSPSA as
\begin{displaymath}
\bar{\mathcal{E}}^{\mbox{\tiny MSPSA}}_T = \sup_{\theta,P,\{f_{i}\}_{i=1}^K} \mathbb{E} \left( \sum_{i=1}^{K} \sum_{t_i=1}^{T_i}2r_{i,t_i} \right).
\end{displaymath}
By (\ref{eq:r_i_t_i}), we bound $\sum_{t_i=1}^{T_i}2 r_{i,t_i} \leq C_0\sqrt{T_i}$ where $C_0 = 4 \max_{i \in \Smsc} C'_i$. Since $T_i$ is smaller than the number of times state $i$ is observed, which is a fraction of $T$ for any $P$, $\bar{\mathcal{E}}^{\mbox{\tiny MSPSA}}_T \leq C\sqrt{T}$ where $C =\sqrt{K}C_0$. Consequently, by (\ref{eq:regret}) and using the upper bound $\bar{\sigma}$ for the singular values of $A_{s_t}$, $\bar{\mathcal{R}}^{\mbox{\tiny MSPSA}}_T \leq \bar{\sigma}^2 \bar{\mathcal{E}}^{\mbox{\tiny MSPSA}}_T \leq C' \sqrt{T}$ where $C' = \bar{\sigma}^2 C$. 
\end{IEEEproof}

According to Theorem~\ref{thm:upper}, the average regret converges to zero with a rate equal or faster than $1/\sqrt{T}$. Hence, it proves that the average performance of MSPSA policy approaches to that of the optimal solution under known model as $T \rightarrow \infty$. However, this does not imply the convergence of the MSPSA policy to the optimal input. The following theorem provides both almost sure and mean square convergence of MSPSA policy to the optimal input.

\begin{theorem}
\label{thm:convergence}
For the quadratic regulation problem,
\begin{IEEEeqnarray}{rCl}
\label{eq:asconv}
\Pr\left(\lim_{T \rightarrow \infty} \|x_T^{\mbox{\tiny MSPSA}}-x^*_{s_{T-1}}\|^2_2 =0 \right) & = & 1,
\end{IEEEeqnarray}
and
\begin{IEEEeqnarray}{rCl}
\label{eq:inputconv}
\lim_{T \rightarrow \infty} \mathbb{E}\left(\|x_T^{\mbox{\tiny MSPSA}}-x^*_{s_{T-1}}\|_2^2\right) & = & 0.
\end{IEEEeqnarray}
\end{theorem}
\begin{IEEEproof}
See Appendix.
\end{IEEEproof}

Theorem~\ref{thm:convergence} shows that, the input generated by MSPSA converges to its optimal value as $T \rightarrow \infty$ for any choice of $\gamma_i>0$. Hence, the condition given in Theorem~\ref{thm:upper} is not necessary for the convergence of MSPSA policy. The intuition is as follows; for any observed recurrent state $i \in \Smsc$, the input of MSPSA converges to its optimal value at time periods when the previous state is $i$ as $T \rightarrow \infty$, and any observed transient state $i \in \Smsc$ will occcur only a finite number of times. Therefore, the input of MSPSA converges to its optimal value as $T \rightarrow \infty$.

\subsection{A Lower Bound on the Growth Rate of Regret}
\label{ssec:lower}

We now show that MSPSA in fact provides the slowest possible regret growth. To this end, we provide a lower bound of regret growth for all decision policies. 

For any policy $\mu$, the estimate of the optimal input and the actual input of the policy may not be the same; for example, the input of MSPSA is a randomly perturbed estimate of the optimal input and not the estimate itself. Hence, let's denote an optimal-input estimate obtained using the past observations corresponding to inputs of policy $\mu$ at time $t$ by $\hat{x}_t^{\mu}$ and the input at time $t$ by $x_t^{\mu}$. We define the worst-case cumulative estimation-MSE as
\begin{displaymath}
\hat{\mathcal{E}}_T^{\mu} \defeq \sup_{\theta,P,\{f_i\}_{i=1}^K} \mathbb{E} \left( \sum_{t=1}^T \|\hat{x}_t^{\mu}-x_{s_{t-1}}^*\|_2^2 \right).
\end{displaymath}

In particular, the following theorem states that the product of the growth rate of the worst-case cumulative input-MSE $\bar{\mathcal{E}}_T^{\mu}$ and the worst-case cumulative estimation-MSE $\hat{\mathcal{E}}_T^{\mu}$ of any sequence $\{\hat{x}_t^{\mu}\}_{t=1}^T$ cannot be lower than $T$ for any policy $\mu$. 

\begin{theorem}
\label{thm:lower}
For any value of $K>1$, there exists a constant $C>0$ such that, for any policy $\mu$, 
\begin{IEEEeqnarray}{rCl}
\label{eq:tradeoffbound}
\hat{\mathcal{E}}_T^{\mu} \bar{\mathcal{E}}_T^{\mu} & \geq & C T.
\end{IEEEeqnarray}
\end{theorem} 
\begin{IEEEproof}
See Appendix.
\end{IEEEproof}

Theorem~\ref{thm:lower} shows the trade-off between exploration (minimizing the estimation error) and exploitation (minimizing the input error). If  the goal is to minimize the cumulative estimation-MSE rather than the regret, than it is possible to find a policy for which $\hat{\mathcal{E}}_T^{\mu}$ grows slower than $\sqrt{T}$ in which case the cumulative input-MSE $\bar{\mathcal{E}}_T^{\mu}$ has to grow faster than $\sqrt{T}$. In fact, if $c_{t_i}$ is set to be constant rather than a decreasing sequence of $t_i$, by following the proof of Theorem~\ref{thm:upper}, it is easy to show that MSPSA's cumulative estimation-MSE grows no faster than $\log{T}$ whereas its regret would grow linearly with $T$.

However, the slowest growth rate of $\bar{\mathcal{E}}_T^{\mu}$ cannot be slower than that of $\hat{\mathcal{E}}_T^{\mu}$ for the optimal choice of the estimate sequence $\{\hat{x}_t^{\mu}\}_{t=1}^T$ (in other words, one can always take the estimate equal to the input, \ie $\hat{x}_t^{\mu}=x_{t}^{\mu}$, in which case $\hat{\mathcal{E}}_T^{\mu}=\bar{\mathcal{E}}_T^{\mu}$). Therefore, the growth rate of the worst-case cumulative input-MSE  $\bar{\mathcal{E}}_T^{\mu}$, and, consequently, the growth rate of the worst-case regret cannot be lower than $\sqrt{T}$ for any policy $\mu$. Therefore, the regret growth rate of MSPSA is the optimal one and achieves $\Omega(\sqrt{T})$ as stated in Theorem~\ref{thm:lower_regret}.

\begin{theorem}
\label{thm:lower_regret}
For any value of $K>1$, there exist some constants $C',C''>0$ such that, for any policy $\mu$, 
\begin{IEEEeqnarray}{rCl}
\label{eq:inputmsebound}
\bar{\mathcal{E}}_T^{\mu} & \geq & C'\sqrt{T},
\end{IEEEeqnarray}
and
\begin{IEEEeqnarray}{rCl}
\label{eq:regretbound}
\bar{\mathcal{R}}_T^{\mu} & \geq & C''\sqrt{T}.
\end{IEEEeqnarray}
\end{theorem} 
\begin{IEEEproof}
We choose the estimate $\hat{x}^{\mu}_t$ equal to the input $x^\mu_t$. Then, by Theorem~\ref{thm:lower}, we have $(\bar{\mathcal{E}}_{T}^{\mu})^2 \geq C T$. As a result, $C'=\sqrt{C}$. Let $\bar{\theta} \in \Theta$ be the parameter satisfying $\mathbb{E} ( \sum_{t=1}^T \|x_t^{\mu}-x_{s_{t-1}}^*\|_2^2 | \bar{\theta} )=\bar{\mathcal{E}}_T^{\mu}$. In Theorem~\ref{thm:lower}, we fixed $p_{i,j}=1/K$ for any $i,j \in \Smsc$. Hence, by (\ref{eq:regret}), $\mathcal{R}_T^{\mu}(\bar{\theta},P) \geq \ubar{\sigma} \bar{\mathcal{E}}_T^{\mu}$ where $\ubar{\sigma}=\min_{\theta \in \Theta} \lambda_{min}(\sum_{j}A_j^{\tra}A_j/K)>0$ by the extreme value theorem. Then, the worst case regret $\bar{\mathcal{R}}_T^{\mu} \geq \mathcal{R}_T^{\mu}(\bar{\theta},P) \geq C'' \sqrt{T}$ where $C'' = \ubar{\sigma}C'$.
\end{IEEEproof}
 
To prove Theorem~\ref{thm:lower}, we consider a hypothetical case in which the decision maker receives additional observations at each period $t$. It is assumed that the additional observations provided to the decision maker are the observation values corresponding to input $x_t^\mu$ from the states that didn't occur at $t$. Since such observations can't increase the growth rate of regret of the optimal policy, we establish a lower bound for this case by showing that it becomes equivalent to a single state case with $m>n$ and using the multivariate van Trees inequality \cite{Gill&etal:95} in a similar way as in  \cite{Keskin&Zeevi:14}. If $K=1$ and $m>n$, the proofs of Theorem~\ref{thm:lower} and Theorem~\ref{thm:lower_regret} lead to the result regarding the single state case given in Corollary~\ref{corollary:lower}. 

\begin{corollary}
\label{corollary:lower}
For $K=1$ and for any value of $m$ and $n$ satisfying $m > n$, there exist some constants $C,C',C''>0$ such that, for any policy $\mu$, inequalities (\ref{eq:tradeoffbound}), (\ref{eq:inputmsebound}), and (\ref{eq:regretbound}) hold.
\end{corollary} 

As mentioned in related work, it has been shown that for $K=1$ and $m=n$ case, the regret growth rate is $\Theta(\log{T})$ \cite{Lai:86}. We show that the characteristics of regret growth changes from $\Theta(\log{T})$ to $\Theta(\sqrt{T})$ for Markov jump system. Additionally, Corollary~\ref{corollary:lower} states that, even in the absence of Markov jump, when system matrix $A$ is not invertible, the best regret growth rate also jumps from $\Theta(\log{T})$ to $\Theta(\sqrt{T})$. These results can be interpreted as a consequence of the fact that the minimum of the cost function for Markov jump system or for single state system with $m>n$ given in (\ref{eq:risk}) is not a root of the cost function as in the case of $K=1$ with $m=n$, and decision maker can't understand how close it is to the minimum just by looking at its observations.

\section{Online Learning for Revenue Maximization}
\label{sec:preliminaries_rm}

The single state setting of the revenue maximization problem has been previously studied by Keskin and Zeevi \cite{Keskin&Zeevi:14}. In this paper, we consider the more general setting where the affine demand parameters can change depending on the state of nature, more precisely the setting where demand parameters follows a Markov jump process. 

As for the quadratic regulation objective, to obtain a regret expression for revenue maximization objective, we first determine the optimal solution of a decision maker who knows the system $(\theta,P)$. Then, we present MSPSA policy for revenue maximization problem and establish its optimality in regret performance and its convergence to the optimal solution.

\subsection{Optimal Solution Under Known Model and Regret} 
By following the same argument as before, under known model, the optimal solution of a decision maker aimed at minimizing the expected cumulative cost given in (\ref{eq:cum_risk}), which is equal to minus expected T-period revenue, is to choose the system input minimizing the respective stage cost given in (\ref{eq:cost_revenue}). By using (\ref{eq:linear}), this stage cost can be written as 
\begin{IEEEeqnarray}{rCl}
\label{eq:risk_revenue}
\mathcal{J}^{\rm}(s_{t-1},x_t) & = & -\sum_{j} p_{s_{t-1},j} x_t^{\tra} \left( A_j x_t+b_j \right). 
\end{IEEEeqnarray}
The optimal input $x_t^*$, which depends only on $(\theta,P)$ and the previous state $s_{t-1}$, is then given by 
\begin{displaymath}
x_{s_{t-1}}^* = - \Big( \sum_j p_{s_{t-1},j}\left(A_j+ A_j^{\tra}\right) \Big)^{-1}\Big(\sum_j p_{s_{t-1},j}b_j\Big),
\end{displaymath}
by dropping the explicit dependency of $x_t^*$ on $(\theta,P)$ in the notation. 

Using the FOC for the optimal input $x_{s_{t-1}}^*$, we obtain the stage regret of a policy $\mu$, \ie
\begin{IEEEeqnarray}{rCl}
r_t^{\mu}(\theta,P) &=& \mathbb{E} \left( \mathcal{J}^{\rm}(s_{t-1},x_t^\mu)-\mathcal{J}^{\rm}(s_{t-1},x_{s_{t-1}}^*) \right) \IEEEnonumber\\
&=& -\mathbb{E} \Big( \big(x_t^{\mu}-x_{s_{t-1}}^* \big)^{\tra}  A_{s_t} \big(x_t^{\mu}-x_{s_{t-1}}^* \big)  \Big). \IEEEnonumber
\end{IEEEeqnarray}
Since $A_{s_t}$ is negative definite, the stage regret is always non-negative. Consequently, the T-period regret is given by
\begin{IEEEeqnarray}{rCl}
\label{eq:cumulative_revenue}
\mathcal{R}_T^{\mu}(\theta,P) & = & -\mathbb{E} \left( \sum_{t=1}^T  \big(x_t^{\mu}-x_{s_{t-1}}^* \big)^{\tra} A_{s_t} \big(x_t^{\mu}-x_{s_{t-1}}^* \big)  \right).\IEEEnonumber\\
\end{IEEEeqnarray}	

\subsection{MSPSA Policy for Revenue Maximization} 

Here, we present the MSPSA policy for revenue maximization objective. The only difference between the two problems considered is their respective stage costs (objectives). Therefore, the only change in MSPSA policy is how the stage costs are calculated to approximate the objective gradient which corresponds to line 11 and 15 of \figurename~\ref{mspsa_for_qr}. In the corresponding steps of MSPSA policy for revenue maximization, that is given in \figurename~\ref{mspsa_for_rm} in details, the stage costs are calculated as minus the observed revenue at that stage.

\begin{figure}
  \begin{boxedalgorithmic}
    \For{t = 1 to T}
   	  \If{$s_{t-1}=i$ is observed}
   	  \If{state $i$ is observed for the first time}
   	  \State Let $\hat{x}_{i,1} \in \Pi$ be an arbitrary vector
   	  \State $t_i \gets 0$
   	  \State $e_i \gets 0$
   	  \EndIf
		\If{$e_i = 0$}
		  \State $t_i \gets t_i+1$
		  \State $x_{t} \gets \hat{x}_{i,t_i}+c_{t_i}\Delta_{t_i}$ 
		  
		  where $c_{t_i}=\gamma'_i/(N'_i+t_i)^{0.25}$ with some positive constant $\gamma'_i$ and a non-negative integer $N'_i$, and $\Delta_{t_i}=[\Delta_{t_i,1},...,\Delta_{t_i,n}]^{\tra}$ with $\Delta_{t_i,j}$'s drawn from an independent and identical distribution that is symmetrical around zero, and satisfies 	$|\Delta_{t_i,j}| \leq \xi_1$ and	$\mathbb{E}(1/\Delta_{t_i,j}^2)\leq \xi_2$  for some positive constants $\xi_1$ and $\xi_2$.
		  	  
		  \State $d_{i,t_i}^{+} \gets -x_{t}^{\tra} y_{t}$ 
		  \State $e_i \gets 1$
		\Else
		  \State $x_{t} \gets \hat{x}_{i,t_i}-c_{t_i}\Delta_{t_i}$
		  \State $d_{i,t_i}^{-} \gets -x_{t}^{\tra} y_{t}$ 
		  \State $e_i \gets 0$
		  \State Update:
			\begin{IEEEeqnarray}{rCl}
				\hat{x}_{i,t_i+1} & \gets & \left(\hat{x}_{i,t_i} - a_{t_i}\left(\frac{d_{i,t_i}^{+}-d_{i,t_i}^{-}}{c_{t_i}}\right)\bar{\Delta}_{t_i}\right)_{\Pi}
			\end{IEEEeqnarray}
			where $(.)_{\Pi}$ denotes the euclidean projection operator onto $\Pi$, $\bar{\Delta}_{t_i} = [1/\Delta_{t_i,1},...,1/\Delta_{t_i,n}]^{\tra}$, and $a_{t_i}=\gamma_i/(N_i+t_i)$ with some positive constant $\gamma_i$ and a non-negative integer $N_i$.
		\EndIf
	  \EndIf 
    \EndFor
  \end{boxedalgorithmic}
\caption{MSPSA for Revenue Maximization} \label{mspsa_for_rm}
\end{figure}

\subsection{MSPSA Performance and Regret Lower Bound} 

To obtain the regret growth rate for MSPSA policy for revenue maximization, we first derive an upper bound on how fast the estimate $\hat{x}_{i,t_i}$ of the optimal input $x_i^*$ converges to its true value as we did for the regulation problem. Lemma~\ref{lemma:regret_revenue} shows that the conditional MSE $e_{i,t_i}$ between the optimal input and its estimate converges to zero with a rate no smaller than the inverse of the square root of the number of times the estimate $\hat{x}_{i,t_i}$ is updated by MSPSA. 

\begin{lemma}
 \label{lemma:regret_revenue}
For any $i \in \Smsc$, if $\gamma_i \geq 1/(8\lambda_{min}(-\sum_j p_{i,j}(A_j+A_j^{\tra})/2))$ then there exists a constant $C_i>0$ satisfying $e_{i,t_i} \leq C_i/\sqrt{t_i}$ for any $(\theta,P,\{f_i\}_{i=1}^K)$.
\end{lemma}
\begin{IEEEproof} See Appendix. \end{IEEEproof}

The condition of Lemma~\ref{lemma:regret_revenue} is slightly different than that of Lemma~\ref{lemma:regret}. The bound on the step size constant $\gamma_i$ depends on the minimum eigenvalue of  $-\sum_j p_{i,j}(A_j+A_j^{\tra})/2$. This difference is due to the choice of a different stage cost. However, the information of a non-trivial lower bound $\ubar{\sigma}$ on the singular values of the system matrices is still sufficient to satisfy this condition.   

Using the result of Lemma~\ref{lemma:regret_revenue}, we prove that MSPSA achieves the regret growth rate of $O(\sqrt{T})$ for revenue maximization objective as given in Theorem~\ref{thm:upper_revenue}, and the input of MSPSA converges to the optimal input both almost surely and in mean square as given in Theorem~\ref{thm:convergence_revenue}.

\begin{theorem}
 \label{thm:upper_revenue}
If  $\gamma_i \geq  1/(8\lambda_{min}(-\sum_j p_{i,j}(A_j+A_j^{\tra})/2))$ for every $i \in \Smsc$, then there exist some positive constants $C$ and $C'$ such that
\begin{displaymath}
\bar{\mathcal{E}}^{\mbox{\tiny MSPSA}}_T \leq C\sqrt{T},
\end{displaymath}
and
\begin{displaymath}
\bar{\mathcal{R}}^{\mbox{\tiny MSPSA}}_T \leq C' \sqrt{T}.
\end{displaymath}
\end{theorem}	
\begin{IEEEproof}
Same as the proof of Theorem~\ref{thm:upper} up to the step that $\bar{\mathcal{E}}^{\mbox{\tiny MSPSA}}_T \leq C\sqrt{T}$ is obtained. Then, by  the regret given in (\ref{eq:cumulative_revenue}) for revenue maximization objective and the fact that $-(A_j+A_j^{\tra})/2$ is positive definite with eigenvalues upper bounded by $\bar{\sigma}$, $\bar{\mathcal{R}}^{\mbox{\tiny MSPSA}}_T \leq \bar{\sigma}\bar{\mathcal{E}}^{\mbox{\tiny MSPSA}}_T \leq C' \sqrt{T}$ where $C' = \bar{\sigma} C$. 
\end{IEEEproof}

\begin{theorem}
\label{thm:convergence_revenue}
For revenue maximization problem,
\begin{displaymath}
\Pr\left(\lim_{T \rightarrow \infty} \|x_T^{\mbox{\tiny MSPSA}}-x^*_{s_{T-1}}\|^2_2 =0 \right) = 1,
\end{displaymath}
and
\begin{displaymath}
\lim_{T \rightarrow \infty} \mathbb{E}\left(\|x_T^{\mbox{\tiny MSPSA}}-x^*_{s_{T-1}}\|_2^2\right) = 0.
\end{displaymath}
\end{theorem}
\begin{IEEEproof}
In the proof of Lemma~\ref{lemma:regret_revenue}, we showed that (\ref{eq:boundforasconv}) holds for any state $i \in \Smsc$. Therefore, the proof follows the proof of Theorem~\ref{thm:convergence}.
\end{IEEEproof}

Previously, for single state setting of this problem, Keskin and Zeevi \cite{Keskin&Zeevi:14} have shown that for any policy $\mu$ the worst case regret growth for this problem cannot be smaller than $\Omega(\sqrt{T})$ and they have shown that $O(\sqrt{T}\log{T})$ is achievable by a semi-myopic policy that they referred to as multivariate constrained iterated least squares (MCILS) policy. Here, we showed that it is possible to achieve the lower bound $\Omega(\sqrt{T})$ given in \cite{Keskin&Zeevi:14} by MSPSA policy for more general problem with Markov jumped demand.

Next, we generalize Keskin and Zeevi's lower bound result to Markov jump case by showing that for any policy $\mu$ and for any state space size $K \geq 1$, the growth rate of worst-case regret is bounded by $\Omega(\sqrt{T})$, and hence MSPSA achieves the optimal rate of $\Theta(\sqrt{T})$. 

\begin{theorem}
\label{thm:lower_revenue}
For any value of $K \geq 1$, there exist some constants $C,C'>0$ such that, for any policy $\mu$,
\begin{IEEEeqnarray}{rCl}
\label{eq:tradeoffbound_revenue}
\hat{\mathcal{E}}_T^{\mu} \bar{\mathcal{E}}_T^{\mu} & \geq & C^2T,
\end{IEEEeqnarray}

\begin{IEEEeqnarray}{rCl}
\label{eq:inputmsebound_revenue}
\bar{\mathcal{E}}_T^{\mu} & \geq & C\sqrt{T},
\end{IEEEeqnarray}
and
\begin{IEEEeqnarray}{rCl}
\label{eq:regretbound_revenue}
\bar{\mathcal{R}}_T^{\mu} & \geq & C'\sqrt{T}.
\end{IEEEeqnarray}
\end{theorem} 
\begin{IEEEproof}
See Appendix.
\end{IEEEproof}

\section{Simulation}
\label{sec:sim}

We present simulation results to illustrate the growth rate of regret and the optimal-input estimate convergence of MSPSA policy both for quadratic regulation and revenue maximization problems. Note that, by these simulation examples, we can only exhibit the performance of "typical" parameters and not the worst-case performance as studied in the theoretical characterization of regret. 
 
For a benchmark comparison, we consider the greedy least square estimate (LSE) method proposed by Anderson and Taylor \cite{Anderson&Taylor:76}. At each period, the greedy LSE determines the input by using the least square estimates of system parameters as if they were the true parameters and projects it onto the set $\Pi$. In order to calculate the initial LSEs of the system parameters, the first samples corresponding to the inputs generated by perturbing the initial input $5\%$ in each direction are taken until the LSE of $\theta$ is computationally tractable. Although, in general, greedy LSE performs well numerically \cite{Anderson&Taylor:76}, it was shown that it can lead to incomplete learning and may not converge with positive probability which causes linear growth in regret \cite{Lai&Robbins:AAM82}.

\begin{figure*}[!t]
\centering
\subfloat[Regret w.r.t. $\sqrt{T}$]{\includegraphics[width=3.5in]{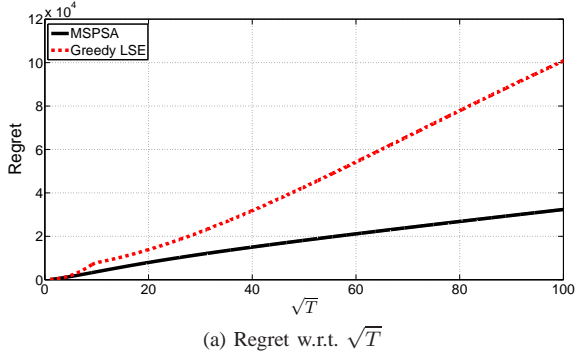}%
\label{sqrtregret_2state}}
\hfil
\subfloat[Optimal-input estimate MSE w.r.t. t]{\includegraphics[width=3.5in]{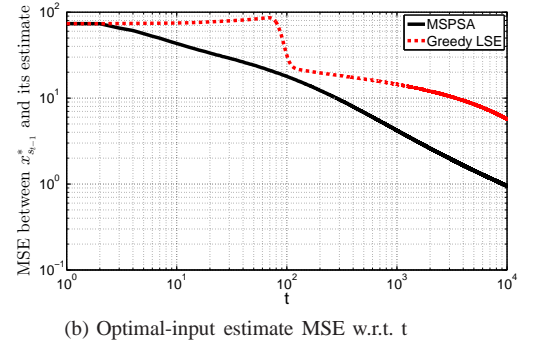}%
\label{priceconv_2state}}
\caption{Average performance of MSPSA and Greedy LSE for quadratic regulation example. The set $\Pi=[1,4]^{24}$, and initial input was set to be the vector of all 4s for both policies. Target value $y^*$ was taken as the vector of all 5s. For each $i \in \Smsc$, $A_i$ and $b_i$ were chosen such that all eigenvalues of $A_i$ belong to the interval $[-1.5,-0.5]$ and the optimal solution associated with each state is contained in $\Pi$. The noise $w_t$ was taken as i.i.d. normal with covariance $0.5^2I_{24}$. The MSPSA parameters were set to be $a_{t_i}=1/(8 \times 0.5)/(t_i+10)$ and $c_{t_i}=1/t_i^{0.25}$; $\Delta_{t_i,j}$'s were drawn from Bernoulli(0.5) with values $\{+1,-1\}$.}
\label{multivariate_target_matching}
\end{figure*}

\begin{figure*}[!t]
\centering 
\subfloat[Regret w.r.t. $\sqrt{T}$]{\psfrag{Cumulative Regret}[c]{\tiny{Regret}}    \includegraphics[width=3.5in]{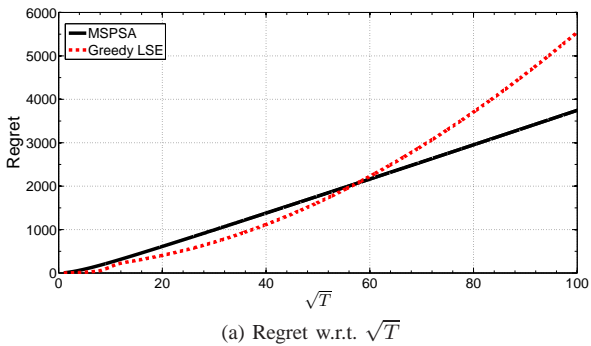} \label{sqrtregret_3state}}
\hfil
\subfloat[Optimal-input estimate MSE w.r.t. t]{\includegraphics[width=3.5in]{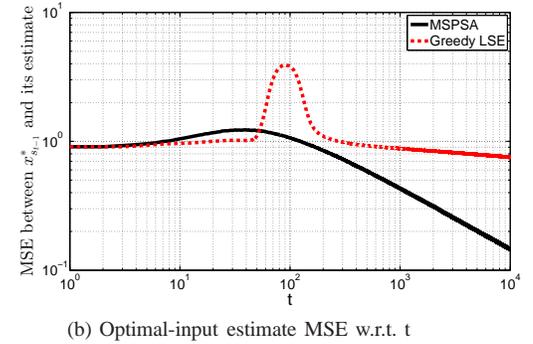}%
\label{priceconv_3state}}
\caption{Average performance of MSPSA and Greedy LSE for revenue maximization example. The set $\Pi=[0.75,2]^{10}$, and initial input was set to be the vector of all 1.375s for both policies. For each $i \in \Smsc$, $A_i$ and $b_i$ were chosen such that all eigenvalues of $A_i$ belong to the interval $[-1.3,-0.3]$ and the optimal solution associated with each state is contained in $\Pi$. The noise $w_t$ was taken as i.i.d. normal with covariance $0.3^2I_{24}$. The MSPSA parameters were set to be $a_{t_i}=1/(8 \times 0.3)/(t_i+10)$ and $c_{t_i}=0.75/t_i^{0.25}$; $\Delta_{t_i,j}$'s were drawn from Bernoulli(0.5) with values $\{+1,-1\}$. }
\label{multivariate_rev_max}
\end{figure*}

\subsection{Numerical Example for Quadratic Regulation Problem}

To illustrate the performance of MSPSA policy for quadratic regulation problem, we consider the problem studied in \cite{Jia&Tong&Zhao:13Allerton}, \ie the problem of an electricity retailer who wants to set hourly electricity prices for the next day to meet its predetermined quantity $y^*$ in demand for each hour. Therefore, the system dimension was set to be $m=n=24$ where each dimension corresponds to an hour of the day. Different than \cite{Jia&Tong&Zhao:13Allerton} where the demand is time-invariant, it is assumed that the demand of its customers changes depending on the state of the day, \eg weather conditions, which follows a Markov jump process. For this example, we considered two states and set the transition probability from any state to the same state to be 0.6 and to the other state to be 0.4. To calculate the average performance, $10^4$ Monte Carlo runs were used. 

\figurename~\ref{multivariate_target_matching} shows the average performance of MSPSA and greedy LSE for this quadratic regulation example. In \figurename~\ref{sqrtregret_2state}, we plot the regret of both policies with respect to square root of the time horizon. We observe that the T-period regret of MSPSA grows linearly with $\sqrt{T}$, which is consistent with the theoretical upper bound. On the other hand, the regret of greedy LSE seems to grow faster than linear. Therefore, we observe that MSPSA outperforms greedy LSE and the difference between the performance of two policies is getting bigger as $T$ increases.
 
\figurename~\ref{priceconv_2state} illustrates how averaged normed squared error between the optimal input and its estimate changes with time in a log-log plot. From Theorem~\ref{thm:convergence}, we know that the optimal-input estimate and thus the input itself converges in mean square for MSPSA. In \figurename~\ref{priceconv_2state}, we observe that the convergence of MSPSA is consistent with this result. Furthermore, the logarithm of the estimation error seems to decrease almost linearly with the logarithm of the time horizon. In other words, MSE seems to converge with a rate equal to $1/\sqrt{t}$. This is reasonable because in Lemma~\ref{lemma:regret}, we show that, for each $i \in \Smsc$, the estimation error decreases with a rate equal or faster than the inverse of the square root of the number of times state $i$ is observed. On the other hand, convergence trend for greedy LSE seems to be much slower and it performs poorly compared with MSPSA's performance. 

\subsection{Numerical Example for Revenue Maximization Problem}

Here, we present an example for revenue maximization problem with system size $n=10$ and 3 different states where the transition probability from any state to the same state was set to be $0.4$ and to any other state to be $0.3$. We used $10^4$ Monte Carlo runs to calculate the average performance of both policies.  

The average performance of MSPSA and greedy LSE for this example with revenue maximization objective is given in \figurename~\ref{multivariate_rev_max}. The regret growth and the convergence of averaged normed squared error between the optimal input and its estimate are illustrated in \figurename~\ref{sqrtregret_3state} and \figurename~\ref{priceconv_3state}, respectively. In both plots, we observe a trend similar to the previous example even though the regret characterizations and the optimal inputs are different due to different objectives. \figurename~\ref{sqrtregret_3state} shows that MSPSA's regret grows linearly with $\sqrt{T}$ whereas greedy LSE's regret grows almost exponentially with $\sqrt{T}$. Therefore, we observe that MSPSA eventually outperforms greedy LSE as $T$ increases even though greedy LSE performs better at the beginning of the time horizon. \figurename~\ref{priceconv_3state} shows that, after a slight increase at the beginning of the time horizon, the logarithm of the estimation error of MSPSA decreases linearly with the logarithm of the time horizon and becomes sufficiently small; whereas the estimation error of Greedy LSE stays almost constant except the spike that is probably due to the poor initial LSEs. Overall, we can say that MSPSA outperforms greedy LSE in both numerical examples.

\section{Conclusion}
We present in this paper an online learning and optimization approach for jump Markov affine models with unknown parameters for two different objectives: (i) quadratic regulation and (ii) revenue maximization. For both objectives, we establish that MSPSA achieves the optimal rate of regret growth $\Theta(\sqrt{T})$. Our results highlight a change of the minimum order of regret growth from $\Theta(\log T)$ of the classical time-invariant affine models to $\Theta(\sqrt{T})$ of the jump Markov affine models for the quadratic regulation objective. On the other hand, we show that introducing a Markov jump process to the system does not change the optimal rate of regret growth for the revenue maximization objective. We also establish the convergence of MSPSA policy to the optimal solution. Our simulation results verify that proposed method MSPSA can outperform the greedy LSE method. 

% Appendix
\appendix

\section*{Proof of Lemma~\ref{lemma:regret}}
Let $\tilde{e}_{i,t_i}=\|\hat{x}_{i,t_i}-x_i^*\|_2^2$. By MSPSA update step given in (\ref{eq:MSPSAupdate_regulation}) and the fact that projection onto $\Pi$ maps a point closer to $x_i^*$, we have,
\begin{IEEEeqnarray}{rCl}
\label{eq:boundt}
\tilde{e}_{i,t_i+1} & \leq & \left \|\hat{x}_{i,t_i} - a_{t_i}\left(\frac{d_{i,t_i}^{+}-d_{i,t_i}^{-}}{c_{t_i}}\right)\bar{\Delta}_{t_i}-x_{i}^*\right\|_2^2 \IEEEnonumber\\
& = & \tilde{e}_{i,t_i} -2 \frac{a_{t_i}}{c_{t_i}}  \left(d_{i,t_i}^{+}-d_{i,t_i}^{-}\right)\left(\hat{x}_{i,t_i}-x_{i}^*\right)^{\tra}\bar{\Delta}_{t_i} \IEEEnonumber\\
&& +\: \frac{a_{t_i}^2}{c_{t_i}^2} \left(d_{i,t_i}^{+}-d_{i,t_i}^{-}\right)^2\bar{\Delta}_{t_i}^{\tra}\bar{\Delta}_{t_i}.%
\end{IEEEeqnarray}

Our goal is to bound $e_{i,t_i+1}= \mathbb{E}(\tilde{e}_{i,t_i+1}|i,t_i)$ by simplifying (\ref{eq:boundt}). By  (\ref{eq:risk}), we obtain,
\begin{IEEEeqnarray}{rCl}
\IEEEeqnarraymulticol{3}{l}{
\mathbb{E}\left( d_{i,t_i}^{+}-d_{i,t_i}^{-} \big|i,t_i,\hat{x}_{i,t_i},\Delta_{t_i} \right)} \IEEEnonumber\\
\qquad \qquad & = & 4c_{t_i} \Delta_{t_i}^{\tra} \sum_{j} p_{i,j} A_j^{\tra}\left(A_j\hat{x}_{i,t_i}+b_j-y^*\right)\IEEEnonumber\\
\qquad \qquad & = & 4c_{t_i} \Delta_{t_i}^{\tra} \Bigl(\sum_{j} p_{i,j} A_j^{\tra} A_j\Bigr)\Bigl(\hat{x}_{i,t_i}-x_{i}^*\Bigr)\IEEEnonumber
\end{IEEEeqnarray}
where last equality is obtained using the FOC for $x_i^*$. Let $\lambda_{min,i}= \lambda_{min}(\sum_j p_{i,j}A_j^{\tra}A_j)$. Using the independence of $\Delta_{t_i,j}$'s, we get,
\begin{IEEEeqnarray}{rCl}
\IEEEeqnarraymulticol{3}{l}{
-\frac{2}{c_{t_i}}\mathbb{E}\left( \left(d_{i,t_i}^{+}-d_{i,t_i}^{-}\right)\left(\hat{x}_{i,t_i}-x_{i}^*\right)^{\tra}\bar{\Delta}_{t_i}\big|i,t_i,\hat{x}_{i,t_i}\right)} \IEEEnonumber\\
\qquad \qquad & = & -8\sum_{j} p_{i,j} \left \|A_j\left(\hat{x}_{i,t_i}-x_{i}^*\right)\right \|_2^2 \IEEEnonumber\\
\qquad \qquad & \leq & -8\lambda_{min,i}\tilde{e}_{i,t_i}. \label{eq:secondterm}%
\end{IEEEeqnarray}

Since $\Pi$ is compact, $\|\hat{x}_{i,t_i} \pm c_{t_i}\Delta_{t_i}\| \leq \bar{x}$ where constant $\bar{x} = (\max_{x \in \Pi}\|x\|)+\gamma_i'\sqrt{n}\xi_1$. For any $j\in \Smsc$, because $b_j$ and singular values of $A_j$ are bounded, $\|A_j(\hat{x}_{i,t_i} \pm c_{t_i}\Delta_{t_i})+b_j-y^*\| \leq C_0$ where constant $C_0=\bar{\sigma}\bar{x}+\bar{b}+\|y^*\|$. By Holder's inequality, we have $\mathbb{E}(\|w_t\|_2^2)\leq m \sigma_w^2$, $\mathbb{E}(\|w_t w_t^{\tra}w_t\|_2)\leq m^2 \sigma_w^3$, and $\mathbb{E}(\|w_t\|_2^4)\leq m^2 \sigma_w^4$. Then, after simplification, we obtain, $\mathbb{E} \left((d_{i,t_i}^\pm)^2|i,t_i,x_{i,t_i}, \Delta_{t_i}\right) \leq C_1$ where $C_1=C_0^4+m^2\sigma_w^4+6C_0^2m\sigma_w^2+4C_0m^2\sigma_w^3$, and 
\begin{IEEEeqnarray}{rCl}
\IEEEeqnarraymulticol{3}{l}{
\mathbb{E}\left(-2d_{i,t_i}^+d_{i,t_i}^- \big|i,t_i,\hat{x}_{i,t_i}, \Delta_{t_i}\right)} \IEEEnonumber\\
\qquad \qquad & \leq & 8c_{t_i}^2\Bigl( \Delta_{t_i}^{\tra} \sum_{j} p_{i,j} A_j^{\tra}\big(A_j\hat{x}_{i,t_i}+b_j-y^*\big)\Bigr)^2 \IEEEnonumber\\
\qquad \qquad & = & 8c_{t_i}^2\bigg( \Delta_{t_i}^{\tra} \Big(\sum_{j} p_{i,j} A_j^{\tra} A_j\Big)\Big(\hat{x}_{i,t_i}-x_{i}^*\Big)\bigg)^2, \IEEEnonumber
\end{IEEEeqnarray}
where last equality is obtained using the FOC of $x_i^*$. Consequently,
\begin{IEEEeqnarray}{rCl}
\label{eq:lastterm}
\mathbb{E} \left( \left(\frac{d_{i,t_i}^{+}-d_{i,t_i}^{-}}{c_{t_i}}\right)^2\bar{\Delta}_{t_i}^{\tra}\bar{\Delta}_{t_i}\Bigg|i,t_i,\hat{x}_{i,t_i}\right) & \leq & C_2\tilde{e}_{i,t_i} + \frac{C_3}{c_{t_i}^2},\IEEEnonumber\\
\end{IEEEeqnarray}
where $C_2=8\max\{2,(1+(n-1)\xi_1^2\xi_2)\}\bar{\sigma}^4$ and $C_3=2C_1n\xi_2$.

Thus, by expressions (\ref{eq:boundt}), (\ref{eq:secondterm}), and (\ref{eq:lastterm}); 
\begin{IEEEeqnarray}{rCl}
\IEEEeqnarraymulticol{3}{l}{
\mathbb{E}(\tilde{e}_{i,t_i+1}|i,t_i,\hat{x}_{i,t_i})} \IEEEnonumber\\
\qquad \qquad & \leq & \big(1-a_{t_i}8\lambda_{min,i} + a_{t_i}^2C_2\big)\tilde{e}_{i,t_i} + a_{t_i}^2\frac{C_3}{c_{t_i}^2}. \label{eq:boundforasconv}
\end{IEEEeqnarray}
Consequently,
\begin{displaymath}
e_{i,t_i+1} \leq \big(1-a_{t_i}8\lambda_{min,i} + a_{t_i}^2C_2\big)e_{i,t_i} + a_{t_i}^2\frac{C_3}{c_{t_i}^2}.
\end{displaymath}
Using this result recursively and since $e^x \geq 1+x$ for all $x \in \Re$, we have,
\begin{IEEEeqnarray}{rCl}
e_{i,t_i+1} & \leq & \Big(\prod_{j=1}^{t_i}\big(1-8a_j\lambda_{min,i}+a_j^2C_2 \big)\Big)e_{i,1} \IEEEnonumber\\
&& +\: \sum_{j=1}^{t_i}\Big(\prod_{l=j+1}^{t_i}\big(1-8a_l\lambda_{min,i}+a_l^2C_2 \big)\Big)a_j^2\frac{C_3}{c_j^2} \IEEEnonumber\\
& \leq & e^{\sum_{j=1}^{t_i} \big(-8a_j\lambda_{min,i}+a_j^2C_2 \big)}e_{i,1} \IEEEnonumber\\
&& +\:  \sum_{j=1}^{t_i} e^{\sum_{l=j+1}^{t_i}\big(-8a_l\lambda_{min,i}+a_l^2C_2 \big)}a_j^2\frac{C_3}{c_j^2}. \IEEEnonumber
\end{IEEEeqnarray}
Since $\gamma_i \geq 1/(8\lambda_{min,i})$ and $e_{i,1} \leq (2\max_{x \in \Pi} \|x\|_2)^2$,
\begin{IEEEeqnarray}{rCl}
e_{i,t_i+1} & \leq & e^{-\log(t_i+1+N_i)+\log(1+N_i)+2\gamma_i^2 C_2}e_{i,1} \IEEEnonumber\\
&& +\: \sum_{j=1}^{t_i}\frac{j+N_i}{t_i+1+N_i} e^{(1+N_i)^{-1}+2\gamma_i^2 C_2 }a_j^2\frac{C_3}{c_j^2} \IEEEnonumber\\
& \leq & \frac{C_i^{'}}{t_i+1+N_i} +\frac{C_i^{''}}{t_{i}+1+N_i} \sum_{j=1}^{t_i}\frac{1}{\sqrt{j+N_i}} \IEEEnonumber\\
& \leq & \frac{C_i}{\sqrt{t_i+1}}, \IEEEnonumber
\end{IEEEeqnarray}
where $C_i^{'} = (1+N_i)\exp(2\gamma_i^2 C_2) 4\max_{x \in \Pi}\|x\|_2^2$, $C_i^{''} = (\gamma_i/\gamma_i')^2 C_3 \exp((1+N_i)^{-1}+2\gamma_i^2 C_2) (1+(\max \{ 0,N_i'-N_i \})^{1/2})$, and $C_i= \max \{C_i^{'},2C_i^{''} \}$.  \hfill \IEEEQED
%%%% New proof
\section*{Proof of Theorem~\ref{thm:convergence}}
In the proof of Lemma~\ref{lemma:regret}, for any state $i \in \Smsc$, we showed that the inequality (\ref{eq:boundforasconv}) holds where $\tilde{e}_{i,t_i}=\|\hat{x}_{i,t_i}-x_i^*\|_2^2$. By Theorem~1~of~Robbins~and~Siegmund~\cite{Robbins:1985}, we know that $\lim_{t_i \rightarrow \infty}\tilde{e}_{i,t_i} < \infty$ exists and $\sum_{t_i=1}^{\infty} 8\lambda_{min,i}a_{t_i}\tilde{e}_{i,t_i} < \infty$ almost surely (a.s.). Since $\sum_{t_i=1}^{\infty} 8\lambda_{min,i}a_{t_i} = \infty$, we obtain that
\begin{displaymath}
\Pr\Big(\lim_{t_i \rightarrow \infty} \tilde{e}_{i,t_i}=0\Big)=1.
\end{displaymath}

Let $1_i(s_{t})$ be the indicator function. Given $s_{t-1}=i$ and $t_i$, $x_t^{\mbox{\tiny MSPSA}}$ is equal to either $\hat{x}_{i,t_i}+c_{t_i} \Delta_{t_i}$ or $\hat{x}_{i,t_i}-c_{t_i} \Delta_{t_i}$. Hence, $ \|x_t^{\mbox{\tiny MSPSA}}-x^*_{i}\|_2^2 \leq 2\tilde{e}_{i,t_i}+2(\gamma_i')^2n\xi_1^2 t_i^{-1/2}$. If state $i$ is recurrent, $\Pr(\lim_{t\rightarrow \infty} t_i < \infty)=0$ because $t_i$ is greater or equal to half of the number of times state $i$ is occurred up to $t$. Therefore, for a recurrent state $i \in \Smsc$,
\begin{IEEEeqnarray}{rCl}
\IEEEeqnarraymulticol{3}{l}{
\Pr\left(\lim_{t \rightarrow \infty} 1_{i}(s_{t-1})\|x_t^{\mbox{\tiny MSPSA}}-x^*_{i}\|_2^2 = 0 \right)} \IEEEnonumber\\
\quad & = & \Pr\left(\lim_{t \rightarrow \infty} 1_{i}(s_{t-1})\|x_t^{\mbox{\tiny MSPSA}}-x^*_{i}\|^2_2 =0 | \lim_{t \rightarrow \infty} t_i = \infty \right) \IEEEnonumber\\
& \geq & \Pr\left(\lim_{t \rightarrow \infty} (2\tilde{e}_{i,t_i}+2(\gamma_i')^2n\xi_1^2 t_i^{-1/2}) = 0 | \lim_{t \rightarrow \infty} t_i = \infty \right) \IEEEnonumber\\
& = & \Pr\Big(\lim_{t_i \rightarrow \infty} \tilde{e}_{i,t_i}=0\Big)=1. \IEEEnonumber
\end{IEEEeqnarray}
So, for a recurrent state $i$ and for any $\epsilon>0$, we have,
\begin{IEEEeqnarray}{rCl}
\label{eq:limrecurrent}
\lim_{t' \rightarrow \infty} \Pr\left( 1_{i}(s_{t-1})\|x_t^{\mbox{\tiny MSPSA}}-x^*_{i}\|^2_2  > \epsilon \mbox{ for some } t \geq t' \right) & = & 0. \IEEEnonumber\\
\end{IEEEeqnarray}

If a state $i \in \Smsc$ is transient, then for any $\epsilon>0$, we have,
\begin{IEEEeqnarray}{rCl}
\IEEEeqnarraymulticol{3}{l}{
\lim_{t' \rightarrow \infty} \Pr\left( 1_{i}(s_{t-1})\|x_t^{\mbox{\tiny MSPSA}}-x^*_{i}\|^2_2  > \epsilon \mbox{ for some } t \geq t' \right)} \IEEEnonumber\\
\qquad \qquad & \leq &\lim_{t' \rightarrow \infty} \Pr\left( s_{t-1}=i \mbox{ for some } t \geq t' \right) = 0, \label{eq:limtransient}
\end{IEEEeqnarray}
where last equality is due to Borel-Cantelli lemma and the fact that $\sum_{t=0}^{\infty}\Pr(s_{t}=i)< \infty$ for a trasient state $i$.

By definition, expression (\ref{eq:asconv}) holds, if and only if, for every $\epsilon >0$, $\lim_{t' \rightarrow \infty} \Pr(\|x_t^{\mbox{\tiny MSPSA}}-x^*_{s_{t-1}}\|^2_2 > \epsilon \mbox{ for some } t \geq t') = 0$. Any state $i \in \Smsc$ is either recurrent or transient. Hence, by (\ref{eq:limrecurrent}) and (\ref{eq:limtransient}), we obtain that, for any $\epsilon >0$,
\begin{IEEEeqnarray}{rCl}
\IEEEeqnarraymulticol{3}{l}{
\lim_{t \rightarrow \infty}\Pr\left(\|x_t^{\mbox{\tiny MSPSA}}-x^*_{s_{t-1}}\|^2_2 > \epsilon \mbox{ for some } t \geq t'\right)} \IEEEnonumber\\
 & \leq &\lim_{t \rightarrow \infty} \sum_{i=1}^K \Pr\left( 1_{i}(s_{t-1})\|x_t^{\mbox{\tiny MSPSA}}-x^*_{i}\|^2_2  > \epsilon \mbox{ for some } t \geq t' \right) \IEEEnonumber\\
& = & 0. \IEEEnonumber
\end{IEEEeqnarray}

Since (\ref{eq:asconv}) holds and $\|x_T^{\mbox{\tiny MSPSA}}-x^*_{i}\|_2^2 \leq C_0$ where $C_0= (2\max_{x \in \Pi}\|x\|+\gamma'_i\sqrt{n}\xi_1)^2$, by Lebesgue's dominated convergence theorem,
\begin{displaymath}
\lim_{t \rightarrow \infty} \mathbb{E}\left(\|x_t^{\mbox{\tiny MSPSA}}-x^*_{s_{t-1}}\|_2^2\right)=0.
\end{displaymath}
\hfill \IEEEQED

%%%% New proof
\section*{Proof of Theorem~\ref{thm:lower}}
Let the transition probability from any state to any other state be $1/K$. Without loss of generality, take $y^*=0$. Let $w_t$ be i.i.d. with distribution $N(\mathbf{0}_m,\sigma_w^2 I_{m})$ which is independent of the state.

Because additional observations can't increase the growth rate of regret for an optimal policy, we assume that the decision maker receives the observation values corresponding to the input $x_t^{\mu}$ from all other states that didn't occur at time $t$ as additional observations at time $t$. Hence, at each $t$, the decision maker gets observations from the affine functions of all states for input $x_t^{\mu}$.
 Let's define $A$, $b$, and $w_t$ as
 \begin{displaymath}
 A = \begin{bmatrix} A_1 \\ \vdots \\ A_K \end{bmatrix} \quad b = \begin{bmatrix} b_1 \\ \vdots \\ b_K \end{bmatrix} 
 \quad  w_t = \begin{bmatrix} w_t^{(1)} \\ \vdots \\ w_t^{(K)} \end{bmatrix} 
 \end{displaymath}
where $w_t^{(i)}$ denotes the system noise of observation from state $i$. Now, for any policy $\mu$, we can express the observation vector at $t$ as
\begin{displaymath}
y_{t}^{\mu} =  Ax_t^{\mu}+b+w_t.
\end{displaymath}

Observe that FOC for the optimal input $x_{s_{t-1}}^*$ at time $t$ obtained from minimizing (\ref{eq:risk}) is the same for any state $s_{t-1} \in \Smsc$ for our fixed choice of $P$. Hence, we drop the dependence on the previous state $s_{t-1}$ along with $P$ and denote it as $x^*(\theta)$, \ie $x^*_{s_{t-1}}=x^*(\theta)$, to express the dependence on $\theta$. With the new notation, FOC can be expressed as 
\begin{displaymath}
 \nu=A^{\tra}(Ax^*(\theta)+b) = 0. 
\end{displaymath}
Consequently, the optimal price given in (\ref{eq:optprice}) becomes
\begin{displaymath}
x^*(\theta) = -(A^{\tra}A)^{-1}A^{\tra}b.
\end{displaymath}

Let's express $\theta_k$ as $\theta_k = [b_{k,1},a_{k,1},...,b_{k,m},a_{k,m}]^{\tra}$ where $b_{k,i}$ is the $i$th entry of $b_k$ and  $a_{k,i}$ is the $i$th row vector of $A_k$. We fix a compact rectangle  $\Theta \subset \Re^{K \times m \times (n+1)}$ such that, for any $\theta \in \Theta$, $x^*(\theta)$ is contained in $\Pi$ and $A_k$ is full column rank for all $k \in \Smsc$. \footnote{The existence of $\Theta$ can be shown by the continuity of $x^*(\theta)$ on a compact rectangle $\Theta'$ which satisfies $A_k$ to be full column rank for all $k \in \Smsc$ and for any $\theta \in \Theta'$, and contains a fixed point $\theta'$ in its interior for which $x^*(\theta')$ is in the interior of $\Pi$. The existance of $\Theta'$ can be shown by using the continuity of the determinant of $A_k^{\tra}A_k$ for each $k \in \Smsc$ at the fixed point $\theta'$.} Since $P$ and $\{f_i\}_{i=1}^K$ are already fixed, our goal is to obtain the performance of the worst-case system parameter $\theta$ that is chosen from the set $\Theta$.

Applying implicit function theorem on $\nu$ gives
\begin{IEEEeqnarray}{rCl}
\label{eq:diffopt}
\frac{\partial x^*(\theta)}{\partial \theta} = -\left(\frac{\partial \nu}{\partial x^*(\theta)}\right)^{-1} \frac{\partial \nu}{\partial \theta} = - \left(A^{\tra}A\right)^{-1}  \frac{\partial \nu}{\partial \theta}, 
\end{IEEEeqnarray}
and by calculus, we have,
\begin{IEEEeqnarray}{rCl}
\label{eq:diffFOC}
\left(\frac{\partial \nu}{\partial \theta}\right)^{\tra} & = &  \left(Ax^*(\theta)+b \right) \otimes \begin{bmatrix} \mathbf{0}_n^{\tra} \\ \mathbb{I}_{n} \end{bmatrix} + A \otimes \begin{bmatrix} 1 \\ x^*(\theta)\end{bmatrix}.
\end{IEEEeqnarray}

Let $M = Km$. Density of the output vector up to time $t$ given the parameter vector $\theta$ and input vector $X^t$ can be written as 
\begin{displaymath}
g(Y^t|X^t,\theta) = \prod_{i=1}^t \frac{\exp\left(-\|y_{i}^\mu-b- Ax_i^\mu\|_2^2 /(2\sigma_w^2)\right)}{(2\pi\sigma_w^2)^{M/2}} .
\end{displaymath}
By writing the joint distribution as a product of conditionals and by the conditional independence of the input for any policy $\mu$ from the parameter $\theta$ given the information history vector $I_{t-1}$, we get,
\begin{displaymath}
\frac{\partial \log{ g(Y^t,X^t |\theta)}}{\partial \theta} = \frac{\partial \log{ g(Y^t|X^t,\theta)}}{\partial \theta} = \frac{1}{\sigma_w^2} \sum_{i=1}^t w_i \otimes \begin{bmatrix}1\\x_i^\mu\end{bmatrix}. 
\end{displaymath}
By using the mixed product property $(A \otimes B)(C \otimes D) = AC \otimes BD$ and the independence of $w_i$, we obtain the fisher information for $g$ as
\begin{IEEEeqnarray}{rCl}
I_t^{\mu}(\theta) & = & \mathbb{E}\left(\frac{\partial \log{ g(Y^t|X^t,\theta)}}{\partial \theta}\frac{\partial \log{g(Y^t|X^t,\theta)}}{\partial \theta}^{\tra} \bigg| X^t, \theta \right) \IEEEnonumber\\
& = & \frac{1}{\sigma_w^2} I_{M} \otimes \left( \sum_{i=1}^t\begin{bmatrix}1 \\ x_i^\mu \end{bmatrix}\begin{bmatrix} 1, (x_i^\mu)^{\tra} \end{bmatrix} \right). \label{eq:fisher}
\end{IEEEeqnarray}
Now, we choose a prior distribution $\lambda$ as an absolutely continuous density on $\Theta$ taking positive values in the interior of $\Theta$ and zero on its boundary. We choose $A$ and $b$ to be independently distributed with distributions $\lambda_{A}$ and $\lambda_b$, respectively, so that $\lambda = \lambda_A \lambda_b $.  Take $ C(\theta)= b^{\tra} \otimes \begin{bmatrix} -x^*(\theta), I_{n} \end{bmatrix}$. Now, we use the multivariate van Trees inequality \cite{Gill&etal:95} in a similar way in \cite{Keskin&Zeevi:14}. This inequality can be expressed as
\begin{IEEEeqnarray}{rCl}
\label{eq:vantreesineq}
\mathbb{E} \left(\|\hat{x}^{\mu}_t - x^*(\theta)\|_2^2\right) & \geq &  \frac{\left(\mathbb{E} \left(\Tr\left(C(\theta)\frac{\partial x^*(\theta)}{\partial \theta}^{\tra}\right)\right)\right)^2 }{\mathbb{E} \left(\Tr\left(C(\theta)I^{\mu}_{t-1}(\theta) C(\theta)^{\tra}\right)\right)+ \tilde{I}(\lambda)} \IEEEnonumber\\
\end{IEEEeqnarray}
where the expectation operators are also taken over the prior distribution $\lambda$ and $\tilde{I}(\lambda)$ is some constant given $\lambda$, which can be seen as the Fisher information for the distribution $\lambda$.

By (\ref{eq:fisher}), we have,
\begin{IEEEeqnarray}{rCl}
\Tr\left(C(\theta)I^{\mu}_{t-1}(\theta)C(\theta)^{\tra}\right) & = &\frac{b^{\tra} b}{\sigma_w^2} \sum_{i=1}^{t-1}\|x_i^\mu-x^*(\theta)\|_2^2 \IEEEnonumber\\
& \leq & c_0 \sum_{i=1}^{t-1}\|x_i^\mu-x^*(\theta)\|_2^2 \label{eq:denom}
\end{IEEEeqnarray}
where $c_0=(K\bar{b}^2)/\sigma_w^2$. 
 
Let's define $P = I_{M} -A (A^{\tra}A)^{-1}A^{\tra}$. Since $A^{\tra}A$ is symmetric positive definite, by (\ref{eq:diffopt}) and (\ref{eq:diffFOC}), we obtain 
\begin{IEEEeqnarray*}{rCl}
\Tr\left(C(\theta)\frac{\partial x^*(\theta)}{\partial \theta}^{\tra}\right) & = & \Tr \left(-C(\theta)\frac{\partial \nu}{\partial \theta}^{\tra}  (A^{\tra}A)^{-1} \right)\\
& = & - b^{\tra} \left(Ax^*(\theta)+b\right) \Tr\left(\left(A^{\tra} A\right)^{-1}\right)\\ 
& = & - b^{\tra} P b \Tr\left((A^{\tra}A)^{-1}\right).\\
\end{IEEEeqnarray*}

By singular value decomposition (SVD) of $A$, observe that $P=U D(\mathbf{0}_n^{\tra}, \mathbf{1}_{M-n}^{\tra}) U^{\tra}$ where $U \in \Re^{M \times M}$ is an orthogonal matrix, and $D(d_1,...,d_{M})$ denotes a diagonal matrix with diagonal entries $d_1,...,d_{M}$. Hence, $P$ is symmetric positive semidefinite. Also observe that $\Tr((A^{\tra} A)^{-1}) \geq n/(K\bar{\sigma}^2)$. Then, we can bound the numerator term,
\begin{IEEEeqnarray}{rCl}
\label{eq:num}
\left(\mathbb{E} \left(\Tr\left(\mathbb{C}(\theta)\frac{\partial x^*(\theta)}{\partial \theta}^{\tra}\right)\right)\right)^2 & \geq &  \frac{n^2}{K^2\bar{\sigma}^4} \left(\mathbb{E}\left(b^{\tra}P b\right)\right)^2.
\end{IEEEeqnarray}

Observe that $\bar{P}= \mathbb{E}(P)$ is symmetric positive semidefinite and nonzero for $K>1$ (or $K=1$ and $m>n$) since $\Tr(\bar{P})=\mathbb{E}(\Tr(P))=Km-n$. Hence, there exists some direction $z \in \Re^M$ such that $z^{\tra}\bar{P}z > 0$, and, consequently, there exists some distribution $\lambda_b$ such that $\mathbb{E}(b)^{\tra}\bar{P}\mathbb{E}(b)>0$. More specifically, if $\mathbb{E}(b)^{\tra}\bar{P}\mathbb{E}(b)=0$ for some choice of $\lambda_b$, we can change that choice of $\lambda_b$ to shift the mean of $b$ slightly in the direction of $z$, and have $\mathbb{E}(b)^{\tra}\bar{P}\mathbb{E}(b)>0$. By independence of $b$ and $P$,
\begin{IEEEeqnarray}{rCl}
\mathbb{E}\left(b^{\tra} P b\right) & = & \mathbb{E}(b)^{\tra} \bar{P} \mathbb{E}(b) +  \mathbb{E}\left( \left(b-\mathbb{E}(b)\right)^{\tra} P \left(b-\mathbb{E}(b)\right)\right) \IEEEnonumber\\
& \geq & \mathbb{E}(b)^{\tra} \bar{P} \mathbb{E}(b). \label{eq:numineq}
\end{IEEEeqnarray}
Hence, by expressions~(\ref{eq:vantreesineq}), (\ref{eq:denom}), (\ref{eq:num}), and (\ref{eq:numineq}); 
\begin{IEEEeqnarray}{rCl}
\sum_{t=2}^T \mathbb{E}\left(\|\hat{x}^{\mu}_t - x^*(\theta)\|_2^2\right) 
& \geq & \sum_{t=2}^T \frac{c_1} {\mathbb{E} (\sum_{i=1}^{t-1}\|x_i^\mu-x^*(\theta)\|_2^2) + c_2} \IEEEnonumber\\
& \geq & \sum_{t=2}^T\frac{c_1} {\mathbb{E} (\sum_{i=1}^{T}\|x_i^\mu-x^*(\theta)\|_2^2) + c_2}, \IEEEnonumber\\*
\label{eq:vantrees}
\end{IEEEeqnarray}
where $c_1 = n^2(\mathbb{E}(b)^{\tra} \bar{P} \mathbb{E}(b))^2/(K^2\bar{\sigma}^4c_0)$ and $c_2 = \tilde{I}(\lambda)/c_0$. 

Since $\bar{\mathcal{E}}_T^{\mu} \geq \mathbb{E} (\sum_{i=1}^{T}\|x_i^\mu-x^*(\theta)\|_2^2)$, by (\ref{eq:vantrees}), we have,
\begin{displaymath}
\hat{\mathcal{E}}_T^{\mu} \geq \frac{c_1(T-1)}{\bar{\mathcal{E}}_T^{\mu}+c_2} \geq \frac{c_1(T-1)}{(1+c_2/\bar{\mathcal{E}}_1^\mu)\bar{\mathcal{E}}_T^{\mu}}. 
\end{displaymath}
Let $x_k^*(\theta)$ denote the $k$th entry of $x^*(\theta)$, and, by extreme value theorem, $u_k = \sup_{\theta \in \Theta} x_k^*(\theta)$ and $l_k = \inf_{\theta \in \Theta} x_k^*(\theta)$ are attained. Since $x^*(\theta)$ is not a constant over $\Theta$ (otherwise $\partial x^*(\theta)/\partial \theta$ would be zero for all $\theta \in \Theta$, and left hand side of (\ref{eq:num}) would be zero for any $\lambda$ which is a contradiction), $\max_{k \in \{1,...,n\}} (u_k - l_k) > 0$. For any policy $\mu$, $\bar{\mathcal{E}}_1^\mu= \sup_{\theta \in \Theta} \mathbb{E} (\|x^\mu_1-x^*(\theta)\|_2^2 |\theta) \geq \max_{k \in \{1,...,n\}} ((u_k - l_k)/2)^2 > 0$. Hence, we have, $\hat{\mathcal{E}}_T^{\mu} \geq  (C T)/\bar{\mathcal{E}}_T^\mu$ where $C =c_1/2/\left(1+(4c_2/\max_{k \in \{1,...,n\}} (u_k - l_k)^2) \right) $.
 \hfill \IEEEQED

\section*{Proof of Lemma~\ref{thm:upper_revenue}}
We will follow the steps in Lemma~\ref{lemma:regret} and simplify inequality~(\ref{eq:boundt}). By (\ref{eq:risk_revenue}), we obtain,
\begin{IEEEeqnarray}{rCl}
\IEEEeqnarraymulticol{3}{l}{
\mathbb{E}\left( d_{i,t_i}^{+}-d_{i,t_i}^{-} \big|i,t_i,\hat{x}_{i,t_i},\Delta_{t_i} \right)} \IEEEnonumber\\
\qquad \qquad & = & -2c_{t_i} \Delta_{t_i}^{\tra} \sum_{j} p_{i,j}\left( \left(A_j+A_j^{\tra} \right)\hat{x}_{i,t_i}+b_j\right) \IEEEnonumber\\
\qquad \qquad & = & - 2 c_{t_i} \Delta_{t_i}^{\tra} \Big(\sum_{j} p_{i,j}\left (A_j + A_j^{\tra} \right ) \Big) \Big(\hat{x}_{i,t_i}-x_{i}^*\Big), \IEEEnonumber
\end{IEEEeqnarray}
where last equality is obtained using the FOC for $x_i^*$. Let $\lambda_{min,i} = \lambda_{min}(-\sum_jp_{i,j}(A_j+A_j^{\tra})/2)$. Using the independence of $\Delta_{t_i,j}$'s, we obtain, 
\begin{IEEEeqnarray}{rCl}
\IEEEeqnarraymulticol{3}{l}{
-\frac{2}{c_{t_i}}\mathbb{E}\left( \left(d_{i,t_i}^{+}-d_{i,t_i}^{-}\right)\left(\hat{x}_{i,t_i}-x_{i}^*\right)^{\tra}\bar{\Delta}_{t_i}\big|i,t_i,\hat{x}_{i,t_i}\right)} \IEEEnonumber\\
\qquad & = & 4\big(\hat{x}_{i,t_i}-x_{i}^*\big)^{\tra} \big( \sum_{j} p_{i,j} (A_j + A_j^{\tra}) \big)\big(\hat{x}_{i,t_i}-x_{i}^*\big) \IEEEnonumber\\
 & \leq & -8\lambda_{min,i}\tilde{e}_{i,t_i}. \IEEEnonumber
\end{IEEEeqnarray}

As in Lemma \ref{lemma:regret}, $\|\hat{x}_{i,t_i}\pm c_{t_i}\Delta_{t_i}\| \leq \bar{x}$. Hence, for any $j \in \Smsc$, $(\hat{x}_{i,t_i}\pm c_{t_i}\Delta_{t_i})^{\tra}(A_j(\hat{x}_{i,t_i}\pm c_{t_i}\Delta_{t_i})+b_j) \leq \bar{\sigma}\bar{x}^2+\bar{b}\bar{x}$. Since $\mathbb{E}(\|w_t\|_2^2)\leq n \sigma_w^2$, $\mathbb{E}\left((d_{i,t_i}^\pm)^2|i,t_i,\hat{x}_{i,t_i}, \Delta_{t_i}\right) \leq C_1$ where constant $C_1=(\bar{\sigma}\bar{x}^2+\bar{b}\bar{x})^2+n \sigma_w^2\bar{x}^2$. Consequently,
\begin{IEEEeqnarray}{rCl}
\IEEEeqnarraymulticol{3}{l}{
\mathbb{E} \left( \left(d_{i,t_i}^{+}-d_{i,t_i}^{-}\right)^2\Big|i,t_i,\hat{x}_{i,t_i},\Delta_{t_i}\right)} \IEEEnonumber\\
\quad & \leq & 2c_{t_i}^2 \bigg( \Delta_{t_i}^{\tra} \Big( \sum_{j} p_{i,j}\left(A_j+A_j^{\tra}\right)\Big)\Big(\hat{x}_{i,t_i}-x_i^*\Big)\bigg)^2 +2C_1, \IEEEnonumber
\end{IEEEeqnarray}
and thus, we obtain (\ref{eq:lastterm}) where $C_2=8\max\{2,(1+(n-1)\xi_1^2\xi_2)\}\bar{\sigma}^2$ and $C_3=2C_1n\xi_2$. Therefore, (\ref{eq:boundforasconv}) holds and the rest of the proof is the same as in Lemma~\ref{lemma:regret}. \hfill  \IEEEQED

\section*{Proof of Theorem~\ref{thm:lower_revenue}}
The inequality given in (\ref{eq:tradeoffbound_revenue}) is used to obtain (\ref{eq:inputmsebound_revenue}) and (\ref{eq:regretbound_revenue}) as in Theorem~\ref{thm:lower_regret}. The proof of inequality (\ref{eq:tradeoffbound_revenue}) follows the proof of Theorem~\ref{thm:lower} with some slight modifications to bound the numerator term of the van Trees inequality due to revenue maximization objective. 

The FOC for $x^*(\theta)$ becomes $\nu=\sum_j ((A_j^{\tra}+A_j)x^*(\theta)+b_j) = 0$ and the optimal price $x^*(\theta) =  (\sum_j -(A_j^{\tra}+A_j ) )^{-1} (\sum_j b_j)$. For this problem, the compact rectangle $\Theta$ is such that, for any $\theta \in \Theta$, $x^*(\theta)$ is contained in $\Pi$ and $A_k$ is negative definite for all $k \in \Smsc$.\footnote{The existence of such a set can shown by the same argument as in Theorem~\ref{thm:lower} by using the continuity of the maximum eigenvalue of $A_k$ rather than the determinant.} The implicit function theorem on $\nu$ gives 
\begin{displaymath}
\frac{\partial x^*(\theta)}{\partial \theta} = - \bigg(\sum_j \left(A_j^{\tra}+A_j \right) \bigg)^{-1}  \frac{\partial \nu
}{\partial \theta}
\end{displaymath}
 where 
\begin{displaymath}
\left(\frac{\partial \nu}{\partial \theta}\right)^{\tra}  = \mathbf{1}_K \otimes \left( x^*(\theta) \otimes \begin{bmatrix} \mathbf{0}_n^{\tra} \\ \mathbb{I}_{n} \end{bmatrix} + \mathbb{I}_{n} \otimes \begin{bmatrix} 1 \\ x^*(\theta)\end{bmatrix} \right).
\end{displaymath} 
We take $\lambda_b$ such that $\mathbb{E}( \sum_j b_j) \neq 0$. Consequently, the numerator term of the van Trees Inequality can be bounded as 
\begin{IEEEeqnarray}{rCl}
\IEEEeqnarraymulticol{3}{l}{
\mathbb{E} \left(\Tr\left(\mathbb{C}(\theta)\frac{\partial x^*(\theta)}{\partial \theta}^{\tra}\right)\right)} \IEEEnonumber\\
\quad \qquad & = & \mathbb{E} \bigg( \Big(\sum_j b_j\Big)^{\tra} x^*(\theta)\Tr \Big(\sum_j -\left(A_j^{\tra}+A_j\right) \Big)^{-1} \bigg) \IEEEnonumber\\
& \geq & \frac{n\mathbb{E} \left(\left(\sum_j b_j\right)^{\tra}\left(\sum_j b_j\right)\right)}{(2K\bar{\sigma})^2} \IEEEnonumber\\
& \geq & \frac{n \mathbb{E} \left(\sum_j b_j\right)^{\tra}\mathbb{E}\left(\sum_j b_j\right)}{(2K\bar{\sigma})^2} > 0.\IEEEnonumber
\end{IEEEeqnarray}
 \hfill \IEEEQED
 
\ifCLASSOPTIONcaptionsoff
  \newpage
\fi




\bibliographystyle{IEEEtran}{\bibliography{journal_reference}}

%\begin{thebibliography}{1}
%\end{thebibliography}


%\begin{IEEEbiography}{Sevi Baltaoglu}
%\end{IEEEbiography}


%\begin{IEEEbiographynophoto}{Sevi Baltaoglu}

%\begin{IEEEbiographynophoto}{Lang Tong}

%\begin{IEEEbiographynophoto}{Qing Zhao}

\end{document}